\documentclass[prd,a4,superscriptaddress,nofootinbib]{revtex4}
\usepackage{amsmath,amssymb,epsfig,natbib}
\usepackage{hyperref}
\usepackage{ifthen}

\begin{document}

\def\eprinttmp@#1arXiv:#2 [#3]#4@{
\ifthenelse{\equal{#3}{x}}{\href{http://arxiv.org/abs/#1}{#1}
}{\href{http://arxiv.org/abs/#2}{arXiv:#2} [#3]}}

\renewcommand{\eprint}[1]{\eprinttmp@#1arXiv: [x]@}
\newcommand{\adsurl}[1]{\href{#1}{ADS}}
\renewcommand{\bibinfo}[2]{\ifthenelse{\equal{#1}{isbn}}{%
\href{http://cosmologist.info/ISBN/#2}{#2}}{#2}}

\newcommand{\cln}{{\cal{N}}}
\newcommand{\nHI}{{n_{HI}}}
\newcommand{\clh}{\mathcal{H}}
\newcommand{\ud}{{\rm d}}
\newcommand{\clk}{\mathcal{K}}

\newcommand{\vort}{\varpi}
\newcommand\ba{\begin{eqnarray}}
\newcommand\ea{\end{eqnarray}}
\newcommand\be{\begin{equation}}
\newcommand\ee{\end{equation}}
\newcommand\lagrange{{\cal L}}
\newcommand\cll{{\cal L}}
\newcommand\clx{{\cal X}}
\newcommand\clz{{\cal Z}}
\newcommand\clv{{\cal V}}
\newcommand\clo{{\cal O}}
\newcommand\cla{{\cal A}}
\newcommand{\uD}{{\mathrm{D}}}
\newcommand{\calE}{{\cal E}}
\newcommand{\calB}{{\cal B}}
\newcommand{\curl}{\,\mbox{curl}\,}
\newcommand\del{\nabla}
\newcommand\Tr{{\rm Tr}}
\newcommand\half{{\frac{1}{2}}}
\newcommand\fourth{{1\over 8}}
\newcommand\bibi{\bibitem}
\newcommand{\kf}{\beta}
\newcommand{\kfprod}{\alpha}
\newcommand\calS{{\cal S}}
\renewcommand\H{{\cal H}}
\newcommand\K{{\rm K}}
\newcommand\mK{{\rm mK}}

\newcommand\opacity{\tau_c^{-1}}

\newcommand{\Psil}{\Psi_l}
\newcommand{\bsigma}{{\bar{\sigma}}}
\newcommand{\bI}{\bar{I}}
\newcommand{\bq}{\bar{q}}
\newcommand{\bv}{\bar{v}}
\renewcommand\P{{\cal P}}
\newcommand{\numfrac}[2]{{\textstyle \frac{#1}{#2}}}

\newcommand{\la}{\langle}
\newcommand{\ra}{\rangle}

\newcommand{\fHe}{f{_\text{He}}}
\newcommand{\Omtot}{\Omega_{\mathrm{tot}}}
\newcommand\xx{\mbox{\boldmath $x$}}
\newcommand{\phpr} {\phi'}
\newcommand{\gam}{\gamma_{ij}}
\newcommand{\sqgam}{\sqrt{\gamma}}
\newcommand{\delk}{\Delta+3{\K}}
\newcommand{\dph}{\delta\phi}
\newcommand{\om} {\Omega}
\newcommand{\dom}{\delta^{(3)}\left(\Omega\right)}
\newcommand{\rar}{\rightarrow}
\newcommand{\Rar}{\Rightarrow}
\newcommand\gsim{ \lower .75ex \hbox{$\sim$} \llap{\raise .27ex \hbox{$>$}} }
\newcommand\lsim{ \lower .75ex \hbox{$\sim$} \llap{\raise .27ex \hbox{$<$}} }
\newcommand\bigdot[1] {\stackrel{\mbox{{\huge .}}}{#1}}
\newcommand\bigddot[1] {\stackrel{\mbox{{\huge ..}}}{#1}}
\newcommand{\Mpc}{\text{Mpc}}
\newcommand{\Al}{{A_l}}
\newcommand{\Bl}{{B_l}}
\newcommand{\eAl}{e^\Al}
\newcommand{\ix}{{(i)}}
\newcommand{\ixp}{{(i+1)}}
\renewcommand{\k}{\beta}
% Derivatives
\newcommand{\HD}{\mathrm{D}}

\newcommand{\nonflat}[1]{#1}
\newcommand{\Cgl}{C_{\text{gl}}}
\newcommand{\Cgltwo}{C_{\text{gl},2}}
\newcommand{\clp}{{\cal P}}
\newcommand{\He}{{\rm{He}}}
\newcommand{\Mhz}{{\rm MHz}}
\newcommand{\vx}{{\mathbf{x}}}
\newcommand{\ve}{{\mathbf{e}}}
\newcommand{\vetilde}{\tilde{\mathbf{e}}}
\newcommand{\vv}{{\mathbf{v}}}
\newcommand{\vk}{{\mathbf{k}}}

\newcommand{\vnhat}{{\hat{\mathbf{n}}}}
\newcommand{\vkhat}{{\hat{\mathbf{k}}}}
\newcommand{\tauc}{{\tau_c}}

\newcommand{\vgrad}{{\mathbf{\nabla}}}
\newcommand{\fbarln}{\bar{f}_{,\ln\epsilon}(\epsilon)}
\newcommand{\etaad}{\eta_a}
\title{The 21cm angular-power spectrum from the dark ages}

\author{Antony Lewis}
\homepage{http://cosmologist.info}

 \affiliation{Institute of Astronomy, Madingley Road, Cambridge, CB3 0HA, UK.}

\author{Anthony Challinor}
 \affiliation{Institute of Astronomy, Madingley Road, Cambridge, CB3 0HA, UK.}
 \affiliation{DAMTP, Centre
for Mathematical Sciences, Wilberforce Road, Cambridge CB3 0WA, UK.}

\date{\today}

\begin{abstract}
At redshifts $z \agt 30$ neutral hydrogen gas absorbs CMB radiation at the 21cm spin-flip frequency. In principle this is observable and a high-precision probe of cosmology.  We calculate the linear-theory angular power spectrum of this signal and cross-correlation between redshifts on scales much larger than the line width.
In addition to the well known redshift-distortion and density perturbation sources a full linear analysis gives additional contributions to the power spectrum. On small scales there is a percent-level linear effect due to perturbations in the 21cm optical depth, and perturbed recombination modifies the gas temperature perturbation evolution (and hence spin temperature and 21cm power spectrum). On large scales there are several post-Newtonian and velocity effects; although negligible on small scales, these additional terms can be significant at $l\alt 100$ and can be non-zero even when there is no background signal. We also discuss the linear effect of reionization re-scattering, which damps the entire spectrum and gives a very small polarization signal on large scales.  On small scales we also model the significant non-linear effects of evolution and gravitational lensing. We include full results for numerical calculation and also various approximate analytic results for the power spectrum and evolution of small scale perturbations.
\end{abstract}
\maketitle

\vskip .2in

\section{Introduction}

The cosmic microwave background (CMB) anisotropies have proved to be a valuable source of information about the initial conditions and evolution of the universe. Most current observations measure the CMB temperature and polarization assuming an exactly blackbody spectrum. However by looking at the anisotropies as a function of frequency vastly more information can be obtained. In addition to the signal from secondary scattering in clusters, in principle there is also line-absorption from sources along the line of sight. One of the most interesting of these is line absorption due to the 21cm spin-flip transition in neutral hydrogen, giving a low frequency probe of the gas distribution at redshifts $300 \agt z \agt 30$~\cite{Scott90,Loeb:2003ya,Furlanetto:2006jb}. This is sensitive to perturbations on all scales down to the Baryon Jeans' scale, which is orders of magnitude smaller than the photon-damping scale that limits what can be learnt directly from the CMB temperature. The 21cm absorption signal therefore potentially contains a huge amount of information about small-scale cosmological perturbations.  Since absorption signal from redshift $z$ is observed at wavelength $\lambda = (1+z) \rm{21.106\,cm}$ [$(1+z)\nu=1420.4\,\Mhz$], the signal can also be studied as a function of observed frequency to give tomographic information about the perturbations~\cite{Loeb:2003ya,Kleban:2007jd}. Unfortunately observations at many-metre wavelengths are very challenging (see e.g. Refs.~\cite{Furlanetto:2006jb,Carilli:2007eb}), but make a useful target for next-but-one generation experiments.

The origin of the dark-age absorption signal is as follows. After recombination there is still a small fraction of free electrons. Compton scattering transfers energy between CMB photons and the electrons (and hence the gas), and hence keeps the gas temperature close to the CMB temperature until a redshift of $z\sim 300$. At lower redshifts the coupling becomes ineffective and the gas starts to cool adiabatically. Atomic collisions in the gas drive the atomic energy levels of the gas towards equilibrium with the gas temperature. The spin temperature defines the relative abundance of triplet and singlet hydrogen states, and is driven by collisions towards the gas temperature. Since the gas cools faster than the CMB, the spin temperature is below the CMB temperature, and 21cm CMB photons will have net absorption by the gas. At redshifts $z \alt 300$ an absorption signal may therefore be observable. At redshifts $z\alt 30$ atomic collisions become very rare, and the spin temperature is driven back towards the CMB temperature by interaction with the numerous CMB photons. The absorption signal from the dark ages is therefore limited to $30\alt z \alt 300$. At lower redshifts sources of Lyman-$\alpha$ photons and non-linear effects become important, and again the spin temperature can depart from the CMB temperature, giving a signal in absorption or emission.

In this paper we focus on the absorption signal from $z\agt 30$ where the physics is well understood and much cleaner than the large uncertainties currently surrounding modelling at lower redshifts. We calculate the linear theory angular power spectrum of the 21cm absorption as a function of redshift, including super-horizon scales where post-Newtonian effects may be important. We focus on the angular power spectrum $C_l(z,z')$ as this is what is directly observable. Many aspects of the physics may be much clearer with a reconstruction of the 3-D power spectrum~\cite{Barkana:2004zy}, but converting the observations into such a spectrum is in general non-trivial especially on large scales, and also dependent on assumptions about the cosmology. Since the perturbations should be nearly linear at high redshifts the statistics should be close to Gaussian, and the angular power spectra should encapsulate most of the statistical information in the observation. Our work extends that of Refs.~\cite{Zaldarriaga:2003du,Furlanetto:2006jb,Bharadwaj:2004nr,Loeb:2003ya,Naoz:2005pd,Barkana:2005xu} by including linear terms due to gravitational redshifting, all velocity affects, ionization fraction perturbations, self-absorption, and reionization re-scattering.  Corrections due to these extra terms are generally quite small, though percent-level effects will be very important if high-redshift 21cm is ever going to fulfil its potential for constraining cosmology.  We also estimate the effect of non-linear evolution, which can be important on small scales even at high redshift, and calculate the effect due to gravitational lensing. Although we do not directly consider $z\alt 30$ here, many of our results could easily be adapted to lower redshifts given a model of the Lyman-$\alpha$ sources and non-linear clustering.

The CMB temperature anisotropy, sourced at $z\sim 10^{3}$, sees super-horizon perturbations at $l\alt 100$. For a 21cm signal at $z\sim 50$ the horizon scale is about ten times larger, corresponding to an angular scale $l\alt 10$. One might therefore expect post-Newtonian effects to dominate at $l< 10$. However the small-scale 21cm anisotropy is sourced by hydrogen density (and spin temperature and redshift distortion) fluctuations, which grow rapidly towards smaller scales. The large scale signal is therefore dominated by fluctuations coming from much smaller scales. Since these smaller scales are uncorrelated on large scales, this gives an approximately white-noise 21cm power spectrum on large angular scales. This white-noise signal dominates that from super-horizon scales, so the post-Newtonian corrections are generally below cosmic variance. In addition there are velocity effects, the most important of which is the dipole in the radiation field seen by each hydrogen atom due to its motion with respect to the CMB. This can give non-negligible quantitative corrections to the angular power spectrum at $l\alt 100$.

On small scales Thomson scattering of 21cm by the background reionization damps the entire spectrum and also induces a very small polarization signal. There are also additional 21cm perturbation sources due to perturbations in the 21cm optical depth; for example an overdensity will have a slightly higher optical depth than the background, leading to a few-percent suppression  in the absorption signal. Additional effects arise indirectly; in particular inclusion of ionization fraction perturbations is important for the evolution of gas temperature fluctuations, and can modify the spectrum at all angular scales by a couple of percent.

The approach we adopt is to evolve the Boltzmann equation for the photon distribution function sourced by absorption of 21cm radiation by neutral hydrogen. We restrict our attention to a spatially-flat close-to-Friedmann-Robertson-Walker cold-dark-matter (CDM) universe. Our results apply equally well with adiabatic or isocurvature scalar mode initial conditions, though we only calculate the adiabatic mode spectra explicitly. We note in passing that 21cm observations are potentially an excellent way to probe isocurvature modes, especially the compensated CDM-baryon mode that cannot be constrained from the CMB temperature~\cite{Barkana:2005xu,Gordon:2002gv}. We do not consider the tensor contribution to the intensity power spectrum as the effect is expected to be well below cosmic variance, though we do calculate the tiny tensor-induced polarization signal. We assume no significant particle decay or annihilations, and assume no variation of constants or non-Gaussianity, though these can be well constrained from their effect on the 21cm signal if included~\cite{Shchekinov:2006eb,Furlanetto:2006wp,Cooray:2006km,Pillepich:2006fj,Valdes:2007cu,Khatri:2007yv}.

We start by deriving the Boltzmann equation for the distribution function in linearized General Relativity. The main linear-theory result relating the observable anisotropy on the sky to sources on the absorption surface is given in Eq.~\eqref{result}. On small scales the majority of the terms are negligible, and we give a result accurate on small scales in Eq.~\eqref{Tbsmall}. This contains the usual density and spin-temperature fluctuation and redshift-distortion sources, but with additional few-percent terms due to the non-zero 21cm optical depth that are often neglected. In Section~\ref{sec:cl} we then derive results for the angular power spectrum in terms of linear-theory transfer functions. To actually calculate the power spectrum we need to calculate the sources, so in Section~\ref{sec:perts} we give results for calculating the background and perturbed densities and temperatures. In Section~\ref{sec:approx} we describe the qualitative shape of the power spectrum, and give approximate semi-analytic results for the large and small scale power spectrum. We then quantify the importance of the various effects in Section~\ref{sec:numerics} where we calculate the 21cm intensity power spectrum numerically. The very small large-scale polarization signal is calculated in Section~\ref{sec:pol}. Since the 21cm power spectrum probes small scales, non-linear evolution can in fact be important at the many-percent level even at redshift $z\sim 50$. We give an approximate estimate of the effect in Section~\ref{sec:nonlin}, and also give an accurate result for the lensed power spectrum in Section~\ref{sec:lens}. A full non-linear analysis is beyond the scope of this paper, but we briefly discuss other sources of non-linear power in Section~\ref{sec:othernonlin}. We finally conclude in Section~\ref{sec:conc}. In a series of appendices we give approximate results for the evolution of small-scale baryon perturbations, general-gauge results for numerical calculation, equations for the evolution of the ionization fraction perturbations, useful results for integrating spherical Bessel functions, and third-order perturbation theory results for the non-linear CDM density and velocity power spectra.

The Boltzmann approach to calculating the line emission angular power spectrum that we consider here is related to the  angular power spectrum of source number counts; we discuss this in a companion paper~\cite{Challinor07}.

\section{Boltzmann equation}

At high redshift, perturbations should be close to linear, and we assume there are no sources of Lyman-$\alpha$ photons. During the dark ages the atomic collision time is comparable to the CMB photon interaction time, and for accurate results the full distribution of spin and velocity states must be accounted for~\cite{Hirata:2006bn}. To simplify our analysis we neglect this complication, focussing on new effects that arise from a full linear perturbation analysis when the spin temperature is independent of atomic velocity. The spin temperature during the dark ages in then governed by 21cm-interaction with CMB photons and atomic collisions. Since the collision rates are not known very accurately, and the ionization fraction after recombination is somewhat uncertain, the precision of our calculation is currently limited anyway.
We also assume the background CMB temperature is exactly blackbody, neglecting any effects due to non-21cm distortions, and assume that atomic angular momenta are isotropically distributed.

We employ linearized standard General Relativity, working in the conformal Newtonian gauge with metric
\begin{equation}
ds^2 = a^2(\eta) [(1+2\psi)d\eta^2 - (1-2\phi) \delta_{ij} dx^i dx^j].
\end{equation}
Except when quoting a few results relevant for calculation of numerical answers we use natural units with $c=1$. We take a velocity field $u^a$ to be along $\partial_\eta$ so that
$u^\mu = a^{-1} (1-\psi) \delta^\mu_0$ and $u_\mu = a(1+\psi)\delta_{\mu 0}$.
This velocity field is the zeroth element of an orthonormal tetrad which
we take to be $(X_0)^a = u^a$ and $X_i \equiv a^{-1} (1+\phi) \partial_i$.
Decomposing a photon wavevector $k^a = dx^a / d\lambda$ into a direction $e^a$
and frequency $k \cdot u \equiv \epsilon/a$, we have
\begin{equation}
\frac{d \vx}{d\eta} = (1+\phi + \psi) \ve , \qquad
\frac{d\eta}{d\lambda} = a^{-2} \epsilon (1-\psi),
\end{equation}
where the three-vector $\ve$ comprises the spatial components of
the propagation direction on the spatial triad $X_i$.

The number density of neutral hydrogen atoms is $\nHI=n_0+n_1$, where the density in the ground state (degeneracy 1) is $n_0$, and the density in the upper triplet state (degeneracy 3) is $n_1$. We assume the spin temperature $T_s$ is dependent only on time and position, defined by $n_1/n_0 = 3  e^{-T_\star/T_s}$ where $T_\star \equiv h_p \nu_{21} /k_B\approx 0.068\,\K$ and $\nu_{21}$ is the constant 21cm frequency. In the rest frame of the gas the net number of 21cm photons emitted per unit volume in proper time $\ud\tau_g$ within energy $\ud E$ within solid angle $\ud\Omega$ is
\begin{equation}
\ud n_{21} = \frac{1}{4\pi}\left[(n_1-3 n_0) \cln_\nu +  n_1\right]A_{10}\Phi(E-E_{21})\ud\tau_g \ud E \ud\Omega,
\end{equation}
where $E_{21}$ corresponds to the 21cm frequency and $\cln_\nu$ is the photon phase space density controlling stimulated emission. The line profile $\Phi(E-E_{21})$ is defined so that $\int \ud E \Phi(E-E_{21}) = 1$. The spontaneous emission rate $A_{10} = 2\pi\alpha\nu_{21}^3 h_p^2/(3c^4 m_e^2) \approx 2.869\times 10^{-15} \rm{s}^{-1}$~\cite{Wild52}, corresponding to a spontaneous decay time of $\sim 10^7$ years and CMB photon interaction time $\sim T_\star/(T_\gamma A_{10})$ (about $10^4$ years at $z=30$). The form of the equation follows from considering detailed balance in equilibrium, in which there is no net production of photons at any temperature.

We shall only make an accurate calculation on scales larger than the line width. In this approximation we model the source emission as monochromatic, so that $\Phi(E-E_{21}) = \delta(E-E_{21})$.
%The observed brightness can be integrated over a frequency window function to account for realistic observational bandwidth.
%In the limit of small 21cm optical depth results can also be integrated over a finite line profile, otherwise they are valid only in the approximation that sources are constant over the redshift interval corresponding to the line width.
The thermal line width corresponds to scales with wavenumber of a few hundred $\Mpc^{-1}$~\cite{Hirata:2006bn,Kleban:2007jd}. This unavoidability suppresses observed power on very small scales regardless of the observational bandwidth. On these scales power is also suppressed due to significant baryon pressure at earlier times, as discussed further below. Observational bandwidth can be accounted for simply by integrating our final result over a frequency window function.

%We assume the escape probability is unity: that each line photon does not interact with any more atoms; this is a reasonable approximation since the optical depth to 21cm absorption is only a few percent.
We model the radiation field as a CMB blackbody $\cln_P = (\exp(h_p\nu/k_B T_\gamma)-1)^{-1}$ plus a term due to 21cm emission $\cln_f$. At the temperatures of interest where $T_\star \ll T_\gamma$ a good approximation is $\cln_P(\nu=\nu_{21})  = T_\gamma/T_\star$ so that $\cln_\nu =  T_\gamma/T_\star + \cln_f(\nu=\nu_{21})$. Usually the photon temperature $T_\gamma$ is taken to be isotropic, but here we are interested in corrections and so allow for its angular variation (e.g. due to the dipole in the rest frame of the atom). The gas-frame temperature is given by
\begin{equation}
T_\gamma^{(g)}(\ve,\vx,\eta) = \bar{T}_\gamma(\eta)\left[1 + \Theta(\ve,\vx,\eta) -\ve\cdot\vv(\vx,\eta)\right],
\end{equation}
where $\Theta$ is the temperature perturbation and $\vv$  the gas (baryon) velocity relative to $u^a$ on the $X_i$ triad. Anisotropy in the radiation field may result in an anisotropic distribution of the atomic triplet states, which would significantly complicate our analysis. This will not be an issue if atomic collisions isotropize the distribution rapidly compared to the photon interaction time. However during the dark ages the collision and photon interaction times are actually similar, so this may not be a safe assumption. Nonetheless, because the interaction is parity invariant, odd multipoles of the radiation field will not affect the triplet distribution; in particular the dominant dipole term leaves an isotropic distribution unchanged. Higher order CMB anisotropies will drive an anisotropy in the triplet distribution, however their relative amplitude is $\sim 10^{-4}$ so their contribution is very small. Also the scattering time for a photon from the anisotropic part of the distribution will be longer than the collision time, so the distribution should in any case be randomized effectively. Similar comments apply to the effect of polarized radiation, so our approximation of an isotropic triplet distribution should be accurate.

If the gas 4-velocity is $u_g^a$ the rest frame energy is given by $E^{(g)}= k_a u_g^a = \epsilon(1-\ve\cdot\vv)/a$, and an interval $\ud\lambda$ along the photon path corresponds to a proper time $\ud\tau_g = u_g^a\ud x_a=k_a u_g^a \ud\lambda $. The Boltzmann equation for the evolution of the distribution function $f$ (number density of photons $f \ud\Omega E^2\ud E/c^3 = 2 \cln_f \ud\Omega \nu^2\ud \nu/c^3$) due to 21cm interaction is then
\begin{eqnarray}
\left.\frac{\ud f}{\ud \lambda}\right|_{H} &=& \frac{c^3 E_{21}}{4\pi E_{21}^2}\frac{3 \nHI A_{10}}{3+e^{T_\star/T_s}}\left[ (1-e^{T_\star/T_s})\left(\frac{T_\gamma^{(g)}}{T_\star}+\frac{h_p^3}{2} f \right)+ 1 \right] \delta(k_a u_g^a- E_{21})\nonumber\\
&\approx& \frac{3 c^3 \bar{n}_{HI} A_{10}}{16 \pi E_{21}}\left[1+\Delta_{HI} - \frac{ \bar{T}_\gamma(1+\Theta-\ve\cdot\vv-\Delta_{T_s}+\Delta_{HI})}{\bar{T}_s}
-\frac{h_p^3}{2}\frac{T_\star}{\bar{T}_s}\left\{ (\Delta_{HI} - \Delta_{T_s})\bar{f} + f
%- \ve\cdot\vv \epsilon\partial_\epsilon \bar{f}
\right\}
\right]\nonumber \\
&&\qquad\qquad\times \delta(\epsilon(1-\ve\cdot\vv)/a - E_{21}),
\end{eqnarray}
where $\nHI$ is the number density of neutral hydrogen and we used the good approximation $T_\star \ll T_s$. We defined fractional perturbations in a quantity $X$ as $\Delta_{X} \equiv (\delta X)/\bar{X}$ and denoted background quantities by an over-bar.
Since the baryon pressure is very low, and the ionization fraction in the dark ages is small, we may take $\Delta_{HI} = \Delta_{H} = \Delta_b$, though the baryon perturbation $\Delta_b$ can differ significantly from the CDM perturbation $\Delta_c$.

Although we do not model 21cm emission from reionization in detail here, we do include re-scattering of emission from higher redshift by the background electron density as this affects the power spectrum from the dark ages. For the moment we neglect polarization and discuss this later. The Thomson scattering contribution to the Boltzmann equation is then
\begin{eqnarray}
\left.\frac{\ud f}{\ud \lambda}\right|_{\text{Thomson}} &=&  E^{(g)} n_e \sigma_T\left[ \frac{3}{16\pi}\int \ud \tilde{\Omega}_{\vetilde'} \tilde{f}(E^{(g)},\vetilde')\left[1 + (\vetilde\cdot \vetilde')^2\right] - f(E,\ve)\right] \nonumber \\
&\approx& E^{(g)} n_e \sigma_T\left[\frac{3}{16\pi}\int \ud \Omega_{\ve'} f(E,\ve')\left[1 + (\ve\cdot \ve')^2\right] - \ve \cdot \vv \epsilon \partial_\epsilon
\bar{f} - f(E,\ve)\right] \nonumber\\
&\approx&\frac{\epsilon \bar{n}_e \sigma_T}{a}\left[ F - \epsilon\partial_\epsilon \bar{f} \ve\cdot\vv + \frac{f_2}{10} - f\right],
\label{boltz_exact}
\end{eqnarray}
where $F$ and $f_2$ are the monopole and quadrupole parts of $f$.
In the first line here, the tildes denote quantities in the gas frame evaluated on the Lorentz-boosted tetrad $\tilde{X}_\mu$.

The background equation does not depend on the Thomson scattering term. Defining
\begin{equation}
\bar{\rho}_s \equiv \frac{3 c^3 \bar{n}_{HI} A_{10}}{16\pi E_{21}^2}\left(\frac{\bar{T}_s - \bar{T}_\gamma}{\bar{T}_s}\right),
\end{equation}
the background equation becomes
\begin{equation}
\frac{\partial \bar{f}}{\partial\eta} = a \bar{\rho}_s \delta(\epsilon/a-E_{21}) - \dot{\bar{\tau}}\bar{f},
\label{eq:line5}
\end{equation}
where the background optical depth to 21cm is defined by
\begin{eqnarray}
\bar{\tau}(\eta,\epsilon) &\equiv& \int_0^\eta \ud\eta'\frac{3 a c^3 \bar{n}_{HI} A_{10}}{16 \pi E_{21}^2} \frac{h_p^3 T_\star}{2\bar{T}_s} \delta(\epsilon/a-E_{21})\nonumber \\
&=& \left[\frac{3\lambda_{21}^2 h_p c A_{10} \bar{n}_{HI}}{32\pi k_B \bar{T}_s H}\right]_\epsilon \theta(\eta-\eta_\epsilon) \nonumber\\
&\equiv& \tau_\epsilon \,\theta(\eta-\eta_\epsilon).
\end{eqnarray}
Here $\epsilon = a(\eta_\epsilon) E_{21}$, $\theta(x)$ is the Heaviside function and $A$ denotes the observation point; a subscript $\epsilon$ denotes the quantity is evaluated at time $\eta_\epsilon$ [and additionally position $\vx_A + \vnhat (\eta_A -
\eta_\epsilon)$ for perturbed quantities along a line of sight $\vnhat$].
 The optical depth $\tau_\epsilon$ is quite small, typically $1$--$4\%$ over the epoch of most interest (see Fig.~\ref{Tb} below). The time derivative is given by
\begin{equation}
\dot{\bar{\tau}} \equiv  a\bar{\tau}_s\delta(\epsilon/a-E_{21})   = \tau_\epsilon \delta(\eta-\eta_\epsilon),
\end{equation}
which defines the optical depth source $\bar{\tau}_s(\eta)$ in analogy with $\bar{\rho}_s$.

The background solution is then given by the integral of Eq.~\eqref{eq:line5},
\begin{equation}
\bar{f}(\eta,\epsilon) =  \frac{1-e^{-\bar{\tau}}}{\tau_\epsilon}\left[\frac{a \bar{\rho}_s }{ E_{21}\clh}\right]_\epsilon
 \equiv
\bar{f}(\epsilon) \frac{1-e^{-\bar{\tau}}}{1-e^{-\tau_\epsilon}},
\label{eq:line6}
\end{equation}
where $\clh$ is the conformal Hubble parameter and $\bar{f}(\epsilon)$ is the value of $\bar{f}(\eta,\epsilon)$ at $\eta>\eta_\epsilon$. To first order in $\tau_\epsilon$ one can use $\bar{f}(\eta,\epsilon)= \bar{f}(\epsilon)\theta(\eta-\eta_\epsilon)$, however the full form given above must be used to get results correct to higher order in $\tau_\epsilon$. Predictions for the 21cm power spectrum are often quoted in terms of
the brightness temperature today, given by
%$T_b = \lambda_{\text{obs}}^2 I_\nu/2k_B = E_{\text{obs}} h_p^3 f/2 k_B$.
$T_b = E_{\text{obs}} h_p^3 f/2 k_B$.
%and $I_\nu =  E_{\text{obs}}^3 h_p f/c^2$,
The isotropic brightness today due to background emission is therefore
\begin{equation}
\bar{T}_b(\eta_A,\epsilon) = (1-e^{-\tau_\epsilon}) \left. \frac{\bar{T}_s-\bar{T}_\gamma}{1+z}\right|_\epsilon.
\end{equation}
During the dark ages the spin temperature is below the CMB temperature, so $\bar{T}_b$ is negative corresponding to net absorption.

To calculate the perturbation to the distribution function we define the monopole source
\begin{equation}
\Delta_s \equiv \Delta_{HI} + \frac{\bar{T}_\gamma}{\bar{T}_s-\bar{T}_\gamma}\left( \Delta_{{T}_s} - \Delta_{T_\gamma}\right).
\end{equation}
The total perturbed Boltzmann equation then becomes
\begin{multline}
\frac{\ud f}{\ud \lambda} = E_{21} \left\{\bar{\rho}_s\left[ 1 +\Delta_s - \frac{\bar{T}_\gamma}{\bar{T}_s-\bar{T}_\gamma}\left\{ \ve \cdot\left(\vv_\gamma -\vv\right)+\Theta_{+} \right\}
\right] - \bar{\tau}_s\left[ \left\{ (\Delta_{HI} - \Delta_{T_s})\bar{f} + f  \right\} \right]\right\}
  \delta(\epsilon(1-\ve\cdot\vv)/a - E_{21})
%- \frac{E_{21}}{a}\dot{\bar{\tau}} \left\{ (\Delta_{HI} - \Delta_{T_s})\bar{f} + f  \right\}\\
%+ \frac{E_{21}}{a} \ve\cdot\vv \bar{f} \epsilon\partial_\epsilon \dot{\bar{\tau}}
\\
+\frac{\epsilon \,\bar{n}_e \sigma_T}{a}\left[ F - \epsilon\partial_\epsilon \bar{f} \ve\cdot\vv + \frac{f_2}{10} - f\right].
\label{eq:exactBoltzmann}
\end{multline}
Here $\Theta_+$ denotes the gauge-invariant temperature anisotropy sources with $l\ge 2$, and $\vv_\gamma$ is the velocity (i.e.\ dipole) of the photon distribution.

To solve the perturbed Boltzmann equation we use the time component of the geodesic equation, which reduces to an equation for the evolution of the comoving energy along the line of sight:
\begin{equation}
\frac{\ud \epsilon}{\ud \eta} = - \epsilon
\frac{\ud \psi}{\ud \eta} + \epsilon (\dot{\phi} + \dot{\psi}),
\end{equation}
where overdots denote conformal-time partial derivatives.
We parameterize the distribution function in the Newtonian gauge as
$f(\eta,\vx,\epsilon,\ve)$, in which case the Boltzmann equation becomes
\begin{multline}
\frac{\partial f}{\partial \eta} + \ve \cdot \vgrad f + \epsilon
\partial_\epsilon \bar{f}\left(\dot{\phi}+\dot{\psi}-\frac{\ud \psi}{\ud \eta}
\right) = a\bar{\rho}_s  \left[1+\Delta_s + \psi-\ve\cdot \vv -\frac{\bar{T}_\gamma}{\bar{T}_s-\bar{T}_\gamma}\left\{ \ve \cdot\left(\vv_\gamma -\vv\right)+\Theta_{+} \right\}\right]
\delta(\epsilon/a-E_{21}) \\
- (\Delta_{HI} - \Delta_{T_s}+\psi-\ve\cdot \vv)\dot{\bar{\tau}} \bar{f}
-  \dot{\bar{\tau}} f
-\ve\cdot \vv\left( \dot{\bar{\tau}} \epsilon\,\partial_\epsilon \bar{f} + \epsilon\,\partial_\epsilon \dot{\bar{f}}\right)
%+ \ve\cdot \vv \bar{f} \epsilon\,\partial_\epsilon \dot{\bar{\tau}}
% - a\bar{\rho}_s  \ve\cdot \vv \,\epsilon\,\partial_\epsilon \delta(\epsilon/a-E_{21})
+ \dot{\tau}_c\left[  \epsilon\partial_\epsilon \bar{f} \ve\cdot\vv - \frac{f_2}{10} + f - F\right],
\label{eq:line4}
\end{multline}
where $\tau_c$ is the Thomson scattering optical depth.
Noting that
$\epsilon \partial_\epsilon = \clh_\epsilon^{-1} \partial_{\eta_\epsilon}$, we have from Eq.~\eqref{eq:line6}
\begin{eqnarray}
\epsilon \partial_\epsilon \bar{f}(\eta,\epsilon)
%&=&
%\frac{1}{\clh_\epsilon} \left[\frac{\partial}{\partial \eta}\left(
%\frac{a\bar{\rho}_s }{E_{21} \clh}\right)\right]_\epsilon \theta(\eta-\eta_\epsilon)
%- \left(\frac{a\bar{\rho}_s }{E_{21} \clh^2}\right)_\epsilon
%\delta(\eta-\eta_\epsilon) \nonumber \\
&=& \fbarln \frac{1-e^{-\bar{\tau}}}{1-e^{-\tau_\epsilon}}
- \frac{\bar{f}(\epsilon)} {\clh_\epsilon}\frac{\dot{\bar{\tau}}e^{-\bar{\tau}}}{1-e^{-\tau_\epsilon}},
\label{eq:line7}
\end{eqnarray}
which defines an additional time-independent quantity $\fbarln$.
Substituting into the Boltzmann equation and integrate formally
along the background line of sight gives the final result
\begin{multline}
\delta f(\eta_A,\vx_A,\epsilon,\vnhat) = \\ e^{-\tau_c}\bar{f}(\epsilon) \left[
\Delta_s + \psi + \vnhat \cdot \vv
+\frac{\bar{T}_\gamma}{\bar{T}_s-\bar{T}_\gamma}\left\{ \vnhat \cdot\left(\vv_\gamma -\vv\right)-\Theta_{+} \right\}+ \frac{\tau_\epsilon e^{-\tau_\epsilon}}{\clh_\epsilon(1-e^{-\tau_\epsilon})}\left(-\frac{\ud \psi}
{\ud\eta} + (\dot{\phi}+\dot{\psi}) +  \vnhat \cdot \frac{\ud \vv}{\ud\eta} \right)
\right]_\epsilon  \\
+e^{-\tau_c}\fbarln\left( e^{\tau_c}\psi_A - \psi_\epsilon + \vnhat\cdot
\vv_\epsilon\right) -\fbarln\int_{\eta_\epsilon}^{\eta_A}\ud \eta  e^{-\tau_c}(\dot{\phi}+\dot{\psi})\\
+e^{-\tau_c}\bar{f}(\epsilon)\left( \frac{\tau_\epsilon e^{-\tau_\epsilon}}{1-e^{-\tau_\epsilon}} -1 \right) \left[\Delta_{HI}-\Delta_{T_s} +\psi+\vnhat\cdot\vv \right]_\epsilon
-  \int_{\eta_\epsilon}^{\eta_A} \ud\eta \dot{\tau_c} e^{-\tau_c} \left( \delta f_0 +  \fbarln (\vnhat\cdot\vv -\psi) + \frac{f_2}{10}\right),
\label{result}
\end{multline}
where $\vnhat = -\ve|_A$ and $\epsilon < a_A E_{21}$ (i.e.\ observed energy strictly less than $E_{21}$), and $\delta f_0$ is the monopole perturbation.
We have assumed that the CMB anisotropies can be well observed at higher frequencies, and the first-order perturbation from last scattering subtracted off the 21cm map. The 21cm brightness fluctuation calculated here is then that of the difference map without the blackbody contribution. We have also assumed there is no overlap between reionization and the 21cm absorption.

The first term in Eq.~\eqref{result} is the usual monopole source, with additional terms $\psi + \vnhat \cdot \vv$ reflecting the additional emission due to the difference between proper time in the gas frame and the interval $\ud\eta$ along the line of sight. Then there is the effect from the CMB dipole seen in the gas frame, $\vv_\gamma-\vv$, plus higher multipole contributions to the temperature seen by the source, $\Theta_+$. The remaining terms on the first line
of Eq.~\eqref{result} describe the local effect of gravitational and Doppler
redshifting on the relation between an observed redshift interval $\Delta z$
and the $\Delta\eta$ along the line of sight. The main such redshift-distortion
effect comes from the radial gradient contribution to
$\vnhat\cdot (\ud\vv/\ud\eta)$. The term multiplying $\fbarln$ is just
$\clh \delta \eta$ where $\delta \eta$ is the perturbation to the conformal
time for a fixed (gas-frame) redshift surface; it has the usual Doppler,
Sachs-Wolfe and integrated Sachs-Wolfe contributions familiar from CMB studies. The first term on the third line describes the perturbation to the 21cm optical depth: for small $\tau_\epsilon$ the term is proportional to $\tau_\epsilon/2$, corresponding to 21cm photons on average seeing half of the perturbation along their line of sight.

To understand the way in which reionization enters Eq.~\eqref{result}, consider
reionization approximated by a delta-function visibility function at
time $\eta_{\text{re}}$ with optical depth $\tau_c$.
Dropping self-scattering terms (taking $\tau_\epsilon\rightarrow 0$) we then have
\begin{multline}
\delta f(\eta_A,\vx_A,\epsilon,\vnhat) =\\ e^{-\tau_c}\bar{f}(\epsilon) \left[
\Delta_s + \psi + \vnhat \cdot \vv
+\frac{\bar{T}_\gamma}{\bar{T}_s-\bar{T}_\gamma}\left\{ \vnhat \cdot\left(\vv_\gamma -\vv\right)-\Theta_{+} \right\}+ \frac{1}{\clh}\left(-\frac{\ud \psi}
{\ud\eta} + (\dot{\phi}+\dot{\psi}) + \vnhat \cdot \frac{\ud \vv}{\ud\eta}
\right)\right]_\epsilon \\
+e^{-\tau_c}\fbarln\left(\psi_A - \psi_\epsilon + \vnhat\cdot
\vv_\epsilon - \int_{\eta_\epsilon}^{\eta_A}  \ud\eta (\dot{\phi}+\dot{\psi})
\right)
+ (1-e^{-\tau_c})\fbarln\left(\psi_A - \psi_{\text{re}} + \vnhat\cdot
\vv_{\text{re}} - \int_{\eta_\text{re}}^{\eta_A}  \ud\eta (\dot{\phi}+\dot{\psi})
\right) \\
+ (1-e^{-\tau_c}) \left(\delta f_0(\epsilon) + \frac{1}{10} f_2(\epsilon,\ve)\right)_{\text{re}} .
\label{eq:reion1}
\end{multline}
The first two sets of terms are the $\delta f$ without reionization mulitplied
by the fraction, $e^{-\tau_c}$, of 21cm photons that are not rescattered.
The third set contains the effective $\delta\eta$ for those
photons that do rescatter (i.e.\ only the common part that is accrued after
reionization) weighted by the fraction, $1-e^{-\tau_c}$, that scatter.
Finally, the fourth terms arise from in-scattering at reionization and
represents an average of the source functions on the electrons' 21cm
surface. For perturbation modes with $k (\eta_{\text{re}}-\eta_\epsilon) \gg 1 $,
the dominant contribution to the 21cm monopole at reionization is
$\delta f_0(\epsilon)|_{\text{re}} \approx \fbarln \psi|_{\text{re}}$ since the
source terms on the electrons' 21cm surface average to zero. Using this
in Eq.~(\ref{eq:reion1}), we see that for such modes reionization damps
the 21cm anisotropies by $e^{-\tau_c}$. For modes with
$k (\eta_{\text{re}}-\eta_\epsilon) \ll 1 $, reionization has essentially no effect
since scattering out of the line of sight is balanced by in-scattering.
However, the contribution from such modes (which were necessarily outside
the horizon at $\eta_\epsilon$) to the 21cm anisotropy at any multipole
$l$ is now small since there is considerably more power in modes with larger
$k$ (see Sec.~\ref{sec:approx}).
The net effect is that reionization should suppress the 21cm anisotropies by
$e^{-\tau_c}$ on all scales, unlike the CMB where there is no suppression at large
$l$.

On small scales (well inside the horizon) most of the terms are completely negligible. Defining $r_\tau\equiv \tau_\epsilon e^{-\tau_\epsilon}/(1-e^{-\tau_\epsilon})$ Eq.~\eqref{result} is approximated by
\begin{eqnarray}
\delta f(\eta_A,\vx_A,\epsilon,\vnhat)
&\approx& e^{-\tau_c}\bar{f}(\epsilon) \left[
\Delta_s - \frac{r_\tau}{\clh_\epsilon}\vnhat \cdot \frac{\partial \vv}{\partial\chi}
 +
\left(r_\tau-1\right) \left(\Delta_{HI}-\Delta_{T_s} \right)\right]_\epsilon
\label{fgoodapprox}
\\
&\approx& e^{-\tau_c}\bar{f}(\epsilon) \left[
\Delta_s - \frac{1}{\clh}\vnhat \cdot \frac{\partial \vv}{\partial\chi}
\right]_\epsilon,
\label{fapprox}
\end{eqnarray}
where $\chi\equiv \eta_A-\eta$ is the conformal distance along the line of sight. The result in the first line should be very accurate on small scales. In the second line we made the approximation that $\tau_\epsilon \sim 0$, $r_\tau \sim 1$ to recover the standard approximation that is only accurate to $\clo(\tau_\epsilon)$ (percent-level). Neglecting small photon perturbations, Eq.~\eqref{fgoodapprox} can also be written as an expression for the brightness perturbation today
\begin{equation}
\delta T_b(\eta_A,\vx_A,\epsilon,\vnhat)=  \frac{e^{-\tau_c}}{1+z_\epsilon}\left[ \tau_\epsilon e^{-\tau_\epsilon}(\bar{T}_s-\bar{T}_\gamma)\left(
\Delta_{HI} - \Delta_{T_s} - \frac{1}{\clh_\epsilon}\vnhat \cdot \frac{\partial \vv}{\partial\chi}\right) +(1-e^{-\tau_\epsilon}) \bar{T}_s \Delta_{T_s}\right]_{\epsilon}.
\label{Tbsmall}
\end{equation}
To consider the impact of the additional terms on large scales we next derive an expression for the power spectrum for numerical calculation.

\section{Angular power spectrum}
\label{sec:cl}
For numerical work one can expand into multipoles and harmonics. We use
\begin{eqnarray}
\delta f(\eta,\vx,\epsilon,\ve) &=& \sum_{l\ge 0} \int \frac{\ud^3 \vk}{(2\pi)^{3/2}} (-i)^l (2l+1)F_l(\eta,\epsilon,\vk) P_l(\vkhat\cdot\ve) e^{i\vk\cdot\vx} \nonumber\\
&=&4\pi\sum_{lm} \int \frac{\ud^3 \vk}{(2\pi)^{3/2}} (-i)^l F_l(\eta,\epsilon,\vk) Y_{lm}^*(\vkhat) Y_{lm}(\ve)e^{i\vk\cdot\vx},
\\
\vv_i(\eta,\vx) &=&  \int \frac{\ud^3 \vk}{(2\pi)^{3/2}} (-i)v_i(\eta,\vk) \vkhat e^{i\vk\cdot\vx},
\end{eqnarray}
for the $i$th species,
and similarly for the temperature multipoles, giving
\begin{multline}
F_l(\eta_A,\epsilon,\vk) =
e^{-\tauc}\left\{
\bar{f}(\epsilon) \left[\Delta_s + \psi + \frac{r_\tau\dot{\phi}}{\clh} +\left(r_\tau-1\right) \left(\Delta_{HI}-\Delta_{T_s} +\psi \right)\right]_\epsilon - \fbarln\psi_\epsilon
\right\}j_l(k\chi_\epsilon)
 \\
%\qquad\qquad
 -e^{-\tauc}\left[ \frac{r_\tau\bar{f}(\epsilon)}{\clh}\left(\dot{v} +\clh v - k\psi \right) +   \fbarln v
+\bar{f}(\epsilon) \frac{\bar{T}_\gamma}{\bar{T}_s-\bar{T}_\gamma} \left(v_\gamma-v\right) \right]_\epsilon  j_l'(k\chi_\epsilon)
+ r_\tau e^{-\tauc}\bar{f}(\epsilon) \frac{k v_\epsilon}{\clh}  j_l''(k \chi_\epsilon) \\
-  \int_{\eta_\epsilon}^{\eta_A} \ud\eta \dot{\tauc} e^{-\tauc}
\left[ (F_0 - \fbarln\psi) j_l(k\chi) - \fbarln v j_l'(k\chi) + \frac{F_2}{4}\left\{ 3j_l''(k\chi)+j_l(k\chi)\right\} \right]
\\
-\fbarln\int_{\eta_\epsilon}^{\eta_A} \ud \eta e^{-\tauc}(\dot{\phi}+\dot{\psi})j_l(k\chi)\,
- e^{-\tauc}\bar{f}(\epsilon) \frac{\bar{T}_\gamma}{\bar{T}_s-\bar{T}_\gamma}
%\sum_{l=2}^\infty 2^{-l}(2l+1)\Theta_l\sum_{m=0}^{l/2} \frac{(2l-2m)!}{m! (l-m)!(l-2m)!}\frac{\ud^{l-2m}}{\ud(k\chi_\epsilon)^{l-2m}} j_l(k\chi_\epsilon)
\sum_{l'=2}^\infty (2l'+1)\Theta_{l'} i^{l'} P_{l'}\left(-\frac{i}{k}\frac{\ud}{\ud\chi_\epsilon}\right)j_l(k\chi_\epsilon)
\label{lofs}
\end{multline}
for $l\ge 1$, where a prime denotes a derivative with respect to the argument.
The $\Theta_l$ are the angular moments of the Fourier expansion of
the CMB temperature anisotropy and are defined analogously to $F_l$.
The last term is small, of the order of the $l\ge 2$ CMB temperature fluctuation.
Note that $\dot{v} + \clh v - k\psi$ is zero in the absence of Thomson scattering or baryon pressure effects.
 Equation~(\ref{lofs}) can be integrated over a given frequency window function (determined by the observation) to determine the actual observed power. If desired one can integrate by parts so that the result depends only on $j_l$ and derivatives of the window function and sources. We  perform our numerical calculations this way in the synchronous gauge: the equations in a general gauge are given in Appendix~\ref{numerics}.

%If we do not make the $T_\star\ll T_s$ approximation there is an additional term $\Delta_{T_s} (T_\star/4T_s)$.
 The angular power spectrum is given by
\begin{equation}
C_l(z,z') = 4\pi \int \ud\ln k\,\, \clp_\chi(k) F_l(k,z) F_l(k,z'),
\label{Cl}
\end{equation}
where $\clp_\chi$ is the power spectrum of the primordial curvature perturbation $\chi$ and $F_l(k,z)$ is the distribution function multipole transfer function to redshift $z=a_A E_{21}/\epsilon-1$; i.e. $F_l(k,z) = F_l(\eta_A,E_{21}/(1+z),\vk)$ for unit initial curvature perturbation.

To calculate the sources for the line-of-sight integral we need to compute the perturbations in the spin and gas temperatures, and, for reionization, the evolution of the low multipoles of the distribution function. We consider these next.

\section{Evolution}
\label{sec:perts}

\begin{figure}
\begin{center}
\epsfig{figure=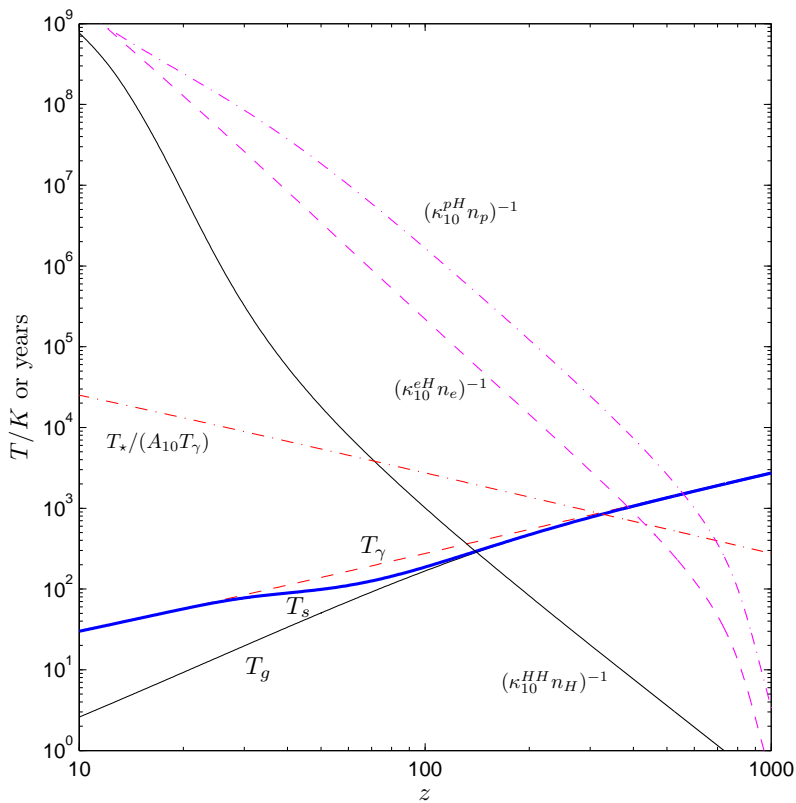,width=10.5cm}
\caption{Evolution of the interaction times for H-H, H-e, H-p and H-photon spin-coupling processes, and how this influences the spin temperature $T_s$ relative to the background CMB and gas temperatures. At high temperatures the H-H collision time is short and collisions couple $T_s$ to the gas temperature $T_g$; at lower redshifts the gas is diffuse and CMB photon interactions drive $T_s$ to the CMB temperature $T_\gamma$. This figure assumes purely linear evolution and no Lyman-alpha coupling; in reality non-linear effects are likely to change the result at $z\alt 30$.
\label{backevolve}
}
\end{center}
\end{figure}

\begin{figure}
\begin{center}
\epsfig{figure=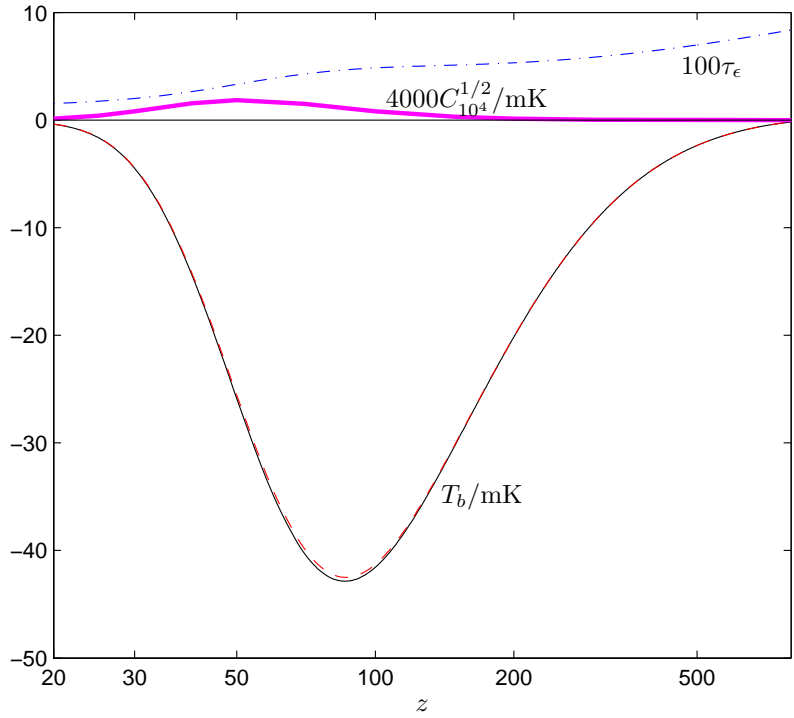,width=10.5cm}
\caption{The background 21cm brightness $T_b$, optical depth $\tau_\epsilon$, and $[l(l+1)C_l/2\pi]^{1/2}$ at $l=10^4$ as a function of source redshift. The dashed line shows the result for $T_b$ neglecting the second term in Eq.~\eqref{Ts_eq} due to the effect of absorption on the ambient blackbody spectrum.
\label{Tb}
}
\end{center}
\end{figure}

\begin{figure}
\begin{center}
\epsfig{figure=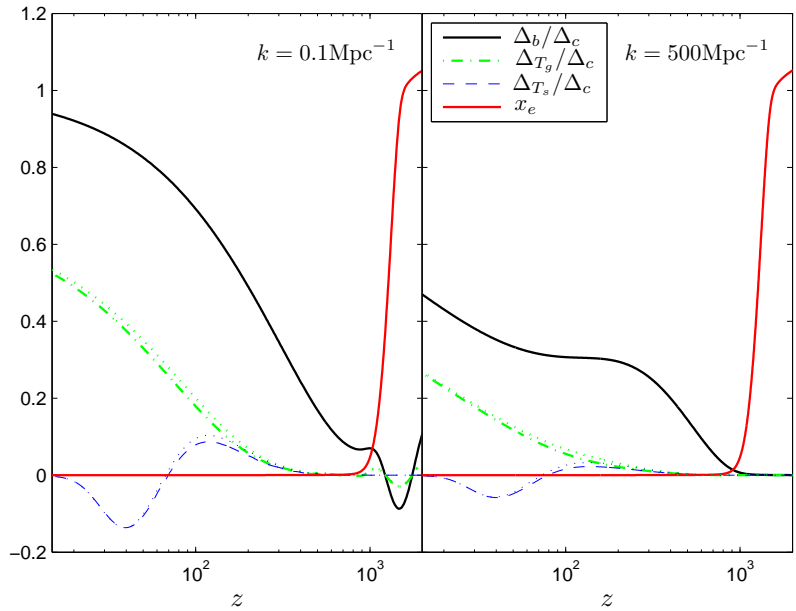,width=10.5cm}
\caption{Evolution of the fractional baryon, matter and spin temperature perturbations as a fraction of the CDM density perturbation. The left figure is for a $k=0.1 \Mpc^{-1}$ mode, the right figure shows the effect of baryon pressure at $k=500 \Mpc^{-1}$.
The dotted lines show the equivalent results neglecting ionization fraction perturbations.
In both cases the baryon perturbation is significantly less than the CDM perturbation at all relevant redshifts. There is no reionization.
% Note that the evolution of the spin temperature differs from the figure in c.f. Ref.~\cite{Bharadwaj:2004nr} by a factor of $\sim 2$.
\label{evolve}
}
\end{center}
\end{figure}

The evolution of the spin temperature is determined from the evolution equation for the states of a fixed number of atoms $N_{HI}=N_0+N_1$ in the gas rest frame. If we crudely assume that recombinations are to the singlet and triplet state in the ratio 1:3 we have
\begin{equation}
\frac{\partial{N}_0 }{\partial\tau}
%+ 3 \clh \bar{n}_0
%- n_0 \vgrad\cdot \vv
= - N_0(C_{01} + 3 A_{10} \cln_{\nu 0}) +  N_1(C_{10} + A_{10}(1+\cln_{\nu 0})) - \frac{\partial x_e}{\partial\tau}\frac{N_{HI} + N_e}{4},
\end{equation}
where $\tau$ is the gas proper time, $\cln_{\nu 0}$ the monopole part
of $\cln_\nu$ in the gas rest-frame integrated over the line profile
(evaluated in Appendix~\ref{app:deltan}),
and the ionization fraction is $x_e\equiv n_e /n_H \equiv n_e/(n_{HI}+n_e)$ (we are neglecting molecular hydrogen and assume all helium is neutral) . Here the collision term is $C_{10} = \kappa_{10}^{HH} n_{HI} + \kappa_{10}^{eH} n_e + \kappa_{10}^{pH}n_p$ and $C_{01} = 3 C_{10} e^{-T_{\star}/T_g}$ where $T_g$ is the gas temperature. The rates $\kappa_{10}^i$ for hydrogen-hydrogen, hydrogen-electron and hydrogen-proton collisional coupling are taken from Refs.~\cite{Furlanetto:2006jb,Furlanetto:2007te} (fit using a cubic splines and 5th order polynomial in the logs), and not known to an accuracy of better than a few percent. As shown in Fig.~\ref{backevolve} the hydrogen-hydrogen term dominates because of the small dark-age ionization fraction ($x_e \sim 2\times 10^{-4}$), and proton-hydrogen rates are suppressed relative to electron-hydrogen rates by a factor of $\sim (m_e/m_p)^{1/2}$ due to the lower proton velocity, except at low redshifts where the proton cross-section is significantly higher~\cite{Furlanetto:2007te}. We include the small correction from electron-hydrogen collisions, but neglect the proton-hydrogen term.

At redshift $z\sim 70$ the total collisional and photon interaction rates are about equal, $C_{10}\sim T_\gamma A_{10}/T_\star$, with $C_{10}\sim 10^{-11}\rm{s}^{-1}$ (corresponding to a collisional coupling time $\sim 4000$ years). At lower redshifts the gas becomes diffuse and collisions become less effective at coupling the spin temperature to the gas temperature.
Defining $\beta_s \equiv 1/T_s$ (etc.), the evolution of the spin temperature is determined by
\begin{equation}
\frac{\partial \beta_s}{\partial\tau}+ \frac{\beta_s}{1-x_e}\frac{\partial x_e}{\partial\tau} = 4\left[ (\beta_g - \beta_s) C_{10} + \beta_\star\left(1 -\beta_s\left[T_\gamma +T_{b0}\right]\right) A_{10}\right],
\label{dbeta}
\end{equation}
where $T_\gamma + T_{b0} = T_* \mathcal{N}_{\nu\,0}$ is the (perturbed) monopole
brightness temperature. For scales large compared to the narrow line
profile, we have
\begin{eqnarray}
T_\gamma + T_{b0} &=& T_\gamma + \frac{1}{2} \hat{\tau}_\eta (T_s - T_\gamma)
\nonumber \\
&=& \bar{T}_\gamma (1+\Delta_{T_\gamma})
+ \frac{\tau_\eta}{2}(\bar{T}_s - \bar{T}_\gamma)\left[1+\Delta_s + \psi
+ \frac{1}{\clh}\left(\dot{\phi} - \frac{1}{3}\vgrad \cdot \vv\right)\right]
\end{eqnarray}
to first-order in the small optical depth $\tau_\eta \equiv \tau_{\epsilon_\eta}$
(see Appendix~\ref{app:deltan}). The quantity
\begin{equation}
\hat{\tau}_\eta \equiv \frac{9 \lambda_{21}^2 h_p c A_{10} n_{HI}}{32\pi
k_B T_s \nabla_a u_g^a},
\label{eq:exactopd}
\end{equation}
is the perturbed optical depth to 21cm, where $\nabla_a u_g^a$ is the volume expansion rate of the gas.
Note that $T_s \hat{\tau}_\eta$ is independent of the (perturbed) spin
temperature.
Since $\tau_\eta$ is small, and the relevant non-perturbative equation
in $\tau_\eta$ cannot be easily solved,
we make this first-order approximation below.
In the epoch before reionization the spin temperature and ionization fraction only vary on Hubble time scales. The coupling time is short compared to the Hubble time, so the spin temperature is determined to very good accuracy by equilibrium, with the left hand side of Eq.~\eqref{dbeta} being zero,
\begin{equation}
T_s \approx T_\gamma\left( \frac{C_{10}T_{\star}/T_\gamma + A_{10}}{C_{10}T_{\star}/T_g + A_{10}}\right)
+\half T_s \hat{\tau}_\eta  A_{10}\left( \frac{1}{C_{10}T_\star/T_g + A_{10}} - \frac{1}{C_{10}T_\star/T_\gamma + A_{10}}\right).
\label{Ts_eq}
\end{equation}
The spin temperature varies between $T_\gamma$ and $T_g$ depending on whether the radiation or collision terms dominate; see Fig.~\ref{backevolve}. The second term in Eq.~\eqref{Ts_eq} due to the finite 21cm optical depth is generally very small, giving a correction to the spin temperature of less than half a percent, and to $T_\gamma-T_s$ of at most about one percent. The small effect on the brightness is shown in Fig.~\ref{Tb}.
This is smaller than the correction due to our assumption of a single velocity-independent spin temperature~\cite{Hirata:2006bn}.

%Perturbations to the spin temperature can be determined from
%\begin{equation}
%\dot{\Delta}_{T_s} = 4 a\left[
%\frac{T_s}{T_g}\left( \Delta_{T_g}-\Delta_{T_s}\right)C_{10} +
%\left(1-\frac{T_s}{T_g}\right)\delta C_{10}  + \left( T_\gamma \Delta_{T_\gamma} - T_s %\Delta_{T_s}\right) \frac{A_{10}}{T_\star} \right] + \frac{\dot{\bar{x}}_e}{(1-\bar{x}_e)^2} %\Delta_{x_e} + \frac{\bar{x}_e}{1-\bar{x_e}} \dot{\Delta}_{x_e}.
%\end{equation}
%(note factor of $g$ seems to be missing in ref.~\cite{Bharadwaj:2004nr} - from numerics looks like an error rather %than a typo.) Evolving this agrees well with the direct equilibrium result

The perturbations to the spin temperature are determined by
\begin{multline}
\Delta_{T_s} - \Delta_{T_\gamma} =  (R_\gamma-R_g)\delta C_{10} + C_{10}(R_g \Delta_{T_g} - R_\gamma\Delta_{T_\gamma})
+\half\tau_\eta A_{10}C_{10}\frac{\bar{T}_g-\bar{T}_\gamma}{T_\star} R_g R_\gamma
\\
\times
\left[ \Delta_{HI} + \psi + \frac{1}{\clh}\left(\dot{\phi}-\frac{1}{3}\vgrad
\cdot \vv\right)
+ \Delta_{C_{10}} + 2\Delta_{T_\gamma}(C_{10}R_\gamma-1) - 2R_\gamma \delta C_{10} + \frac{\bar{T}_\gamma}{\bar{T}_g - \bar{T}_\gamma}\left(\Delta_{T_g}-\Delta_{T_\gamma}\right)\right]
%\nonumber\\
%&=& \left(\Delta_H C_{10} +
%\left[\frac{\ud \ln \kappa_{10}^{HH}}{\ud \ln T_g} (1-\bar{x}_e) \kappa_{10}^{HH} +
%\frac{\ud \ln \kappa_{10}^{eH}}{\ud \ln T_g} \bar{x}_e \kappa_{10}^{eH}\right] \bar{n}_H\Delta_{T_g} +  %(\kappa^{eH}_{10}-\kappa^{HH}_{10})\bar{x}_e \bar{n}_H \Delta_{x_e}\right)(R_\gamma-R_g) \nonumber\\
%&&+ C_{10}(R_g \Delta_{T_g} - R_\gamma\Delta_{T_\gamma}),
\label{delta_spin}
\end{multline}
where $R_i^{-1} \equiv (C_{10} + A_{10} \bar{T}_i/T_\star)$ and
$$
\delta C_{10} = \Delta_H C_{10} +
\left[\frac{\ud \ln \kappa_{10}^{HH}}{\ud \ln T_g} (1-\bar{x}_e) \kappa_{10}^{HH} +
\frac{\ud \ln \kappa_{10}^{eH}}{\ud \ln T_g} \bar{x}_e \kappa_{10}^{eH}\right] \bar{n}_H\Delta_{T_g} +  (\kappa^{eH}_{10}-\kappa^{HH}_{10})\bar{x}_e \bar{n}_H \Delta_{x_e}.
$$
The background ionization fraction $\bar{x}_e$ is taken from RECFAST~\cite{Seager:1999km}. The effect of the optical depth term in Eq.~\eqref{delta_spin} is only about $1\%$ on the angular power spectrum.

\begin{figure}
\begin{center}
\epsfig{figure=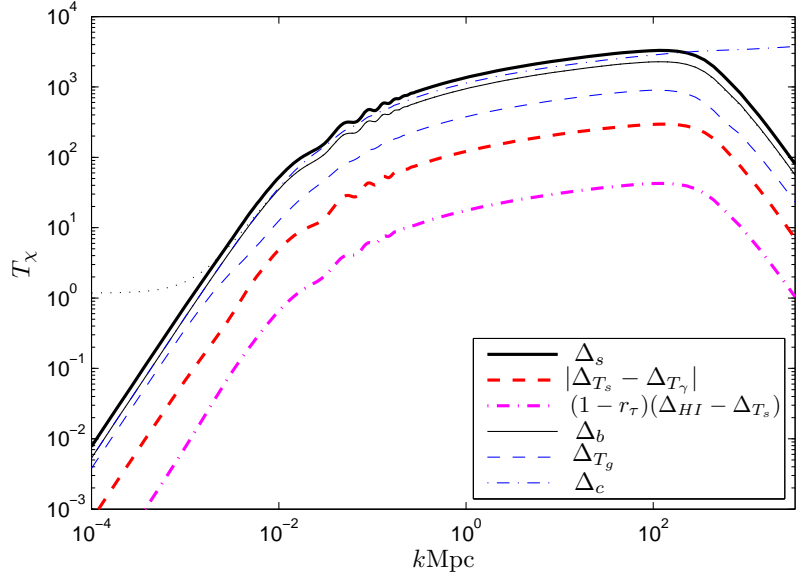,width=10.5cm}
\caption{Transfer function for monopole source at redshift $z=50$ given unit initial curvature perturbation, compared to other relevant perturbations. The perturbations are numerically evaluated in the synchronous gauge. The Newtonian-gauge
fluctuations equal those in the synchronous gauge well inside the horizon ($k \gg 10^{-3}\, \text{Mpc}^{-1}$). On large scales, the Newtonian-gauge $\Delta_b$
flattens out at $\approx 6/5$ times the primordial curvature perturbation, as shown by the dotted curve.
The spin temperature perturbation is negative at $z=50$ (see Fig.~\ref{evolve}).
\label{transfer}
}
\end{center}
\end{figure}

Assuming purely Compton cooling, the background gas temperature evolves according to~\cite{Weymann65,Seager:1999km}
\begin{equation}
\dot{\bar{T}}_g + 2 \clh \bar{T}_g  = -\frac{8 a \sigma_T \bar{\rho}_\gamma \bar{x}_e}{3m_e c (1+f_{\text{He}}+\bar{x}_e)}\left(\bar{T}_g - \bar{T}_\gamma\right),
\end{equation}
where $\sigma_T$ is the Thomson scattering cross-section,
$f_{\text{He}}=n_{\text{He}}/n_H$ and we have ignored the very small effect of
21cm radiation.
The perturbations evolve with
\begin{equation}
\dot{\Delta}_{T_g} =
 2\dot{\phi}  - \frac{2}{3} kv
%-\frac{2}{3}\frac{k}{S}(Z + v)
 -\frac{8 a \sigma_T \bar{\rho}_\gamma \bar{x}_e}{3 m_e c(1+\fHe+\bar{x}_e)}\left[ \left(1-\frac{\bar{T}_\gamma}{\bar{T}_g}\right) \left\{4\Delta_{T_\gamma} + \psi + \frac{\Delta_{x_e}}{1+\bar{x}_e/(1+\fHe)}\right\}
+ \frac{\bar{T}_\gamma}{\bar{T}_g} \left(\Delta_{T_g} - \Delta_{T_\gamma}\right)
\right] ,
\end{equation}
where we neglected helium fraction perturbations. Note that although the direct contribution of $\Delta_{x_e}$ in Eq.~\eqref{delta_spin} is small, the indirect effect on the evolution of $\Delta_{T_g}$ can be significant. The equations for approximately calculating $\Delta_{x_e}$ are given in Appendix~\ref{xepert}. An overdensity has positive $\Delta_{T_g}$ but recombines more fully than the background and hence has negative $\Delta_{x_e}$; the additional effect of the ionization fraction perturbation is therefore to reduce the coupling to the CMB and hence slightly decrease the spin temperature. Typical transfer functions for the perturbations are shown in Fig.~\ref{transfer}
\emph{in the synchronous gauge} that we use for numerical work. The
Newtonian-gauge functions differ on super-horizon scales ($k \alt 10^{-3}\,
\mathrm{Mpc}^{-1}$ at $z=50$). For example, the CDM transfer function
flattens to $\approx 6/5$ on large scales.

Assuming an ideal gas, the gas pressure perturbation $\delta p/\bar{\rho}_g = c_s^2 \Delta_g$ is given by
\begin{equation}
c_s^2 \Delta_g = \frac{k_B \bar{T}_g}{\mu}\left(\Delta_g + \Delta_{T_g}\right)
\approx \frac{k_B \bar{T}_g}{m_p}\left([1+\bar{x}_e](1-Y_\He) + Y_\He m_p/m_\He\right)\left(\Delta_g + \Delta_{T_g}\right),
\label{sound_speed}
\end{equation}
where $\mu$ is the mean particle mass and $Y_\He$ is the mass fraction in helium. This result must be used on scales where the baryon pressure is important~\cite{Naoz:2005pd}. Note that on these scales there may also be additional effects due to CDM decoupling~\cite{Loeb:2005pm} that we neglect here. The evolution of the gas and spin temperature perturbations is shown in Figure~\ref{evolve} for
two different scales, along with the relative evolution of the baryon and CDM perturbations. Again, these are in the synchronous gauge but they are very
similar to the Newtonian-gauge perturbations since both wavelengths are
sub-horizon for the redshift range plotted.

To evaluate the sources for reionization Thomson scattering we need to evolve the multipole equation
\begin{multline}
\dot{F}_l + \frac{k}{2l+1}\left[ (l+1)F_{l+1} - l F_{l-1}\right] + \delta_{l0} \epsilon\partial_\epsilon \bar{f} \dot{\phi}+\frac{\delta_{l1}}{3} \epsilon\partial_\epsilon \bar{f} k\psi =
\\
a \bar{\rho}_s\left[ \delta_{l0}(\Delta_s + \psi) - \frac{\delta_{l1}}{3}\left(\frac{\bar{T}_\gamma}{\bar{T}_s - \bar{T}_\gamma}\right)(v_\gamma-v) - (1-\delta_{l0}-\delta_{l1})\frac{\bar{T}_\gamma}{\bar{T}_s - \bar{T}_\gamma}\Theta_{l} \right]\delta(\epsilon/a-E_{21}) \\
%- \delta_{l1}a\bar{\rho}_s \frac{v}{3} \epsilon\partial_\epsilon \delta(\epsilon/a-E_{21})
-\dot{\tau}_c\left[ (\delta_{l0}-1)F_l - \frac{\delta_{l1}}{3}v \epsilon\partial_\epsilon\bar{f} + \delta_{l2}\frac{F_2}{10}\right]
-\dot{\bar{\tau}} \left[F_l+\delta_{l0}(\Delta_{HI}-\Delta_{T_s}+\psi)\bar{f} \right]
-\delta_{l1} \frac{v}{3}\left(  \partial_\epsilon [\epsilon\dot{\bar{f}}] +\dot{\bar{\tau}}\epsilon\partial_\epsilon\bar{f}\right) .
\label{eq:evolveml}
\end{multline}
Neglecting the self-absorption terms involving $\dot{\bar{\tau}}$ this is straightforward to propagate numerically after integration over a window function in frequency to remove the delta-functions, as discussed further in Appendix~\ref{numerics}. Other perturbed quantities and the photon temperature and polarization multipoles evolve according to standard results as implemented in the numerical codes CMBFAST and CAMB~\cite{Ma:1995ey,Seljak:1996is,Hu:1998mn,Lewis:1999bs}.

\section{Approximate results}
\label{sec:approx}

The general form of the equal-redshift angular power spectra can easily be understood. On super-horizon scales the potential $\phi$ is close to scale invariant and constant, so (from the Poisson equation in the comoving gauge) $\Delta_c \sim k^2/\clh^2\phi$. Hence on entering the horizon ($k\sim \clh$) the matter perturbations are of order of the potential, $\Delta_c \sim\phi$. During radiation domination photon pressure prevents gravitational collapse: the perturbations only grow logarithmically. As a result the spectrum of $\Delta_c$ is approximately scale invariant on sub-horizon scales at matter-radiation equality, though super-horizon scales still have $\Delta_c \sim k^2/\clh^2\phi$. During matter domination the potential remains constant and $\Delta_c$ grows at the same rate independent of $k$ on all scales where baryon pressure can be neglected. The result is that an initially scale-invariant potential gives comoving-gauge CDM perturbations with an amplitude scaling as $k^2$
on large scales and growing logarithmically with $k$ on small scales.
The Newtonian-gauge
CDM perturbation equals that in the comoving-gauge well inside the horizon,
but on larger scales the Newtonian-gauge $\Delta_c \approx -2 \phi$ in matter
domination.

For super-horizon modes, the dominant 21cm sources in the Newtonian gauge
are
\begin{equation}
\delta f(\eta_A,\vx_A,\epsilon,\vnhat) \approx \bar{f}
(\epsilon) \left( \Delta_s + \psi \right) - \fbarln \psi ,
\end{equation}
where we have ignored reionization and the effect of non-zero 21cm optical depth.
The source $\Delta_s$ scales with the hydrogen perturbation $\Delta_{HI}$. After
recombination almost all the atoms are neutral and $\Delta_{HI}=\Delta_b$. On scales above the baryon sound horizon at recombination the baryons fall into the CDM potential wells on a Hubble time scale, so $\Delta_b$ evolves to follow $\Delta_c$.
% with
%\begin{equation}
%\frac{\partial^2}{\partial\eta^2}(\Delta_c-\Delta_b)+ \clh \frac{\partial}{\partial\eta}(\Delta_c-\Delta_b)\approx k^2 c_s^2 %\Delta_b.
%\end{equation}
%Hence $\partial_\eta(\Delta_c-\Delta_b)$ falls off as $1/a$ on scales where $c_s^2 k^2 \ll \clh^2$, even though $\Delta_c$ is %growing as with $a$.
Note that although $\Delta_b$ qualitatively follows $\Delta_c$ well after recombination, the difference can be tens of percent on all sub-horizon scales at high redshift. On very small scales growth of $\Delta_b$ is suppressed once the perturbation reaches pressure support. The Newtonian-gauge 21cm sources
thus have a scale-invariant amplitude on super-horizon scales,
scale as $k^2$ for sub-horizon modes that entered the horizon after
matter-radiation equality,
are growing logarithmically on small scales, and are suppressed on very small scales where $c_s^2 k^2/\clh^2 \agt 1$ at recombination. This behaviour can be seen in Fig.~\ref{transfer}.

On small scales the fractional source fluctuation is of the order of the CDM density perturbation, $\Delta_s \sim (k/\clh)^2\phi$, and the velocity is given by $v \sim (\clh/k)\Delta_c$. For small scales with $k\gg \clh$ the line-of-sight result in Eq.~\eqref{lofs} is therefore dominated by the monopole and redshift distortions effects, giving the usual approximation for the 21cm source when we neglect self-absorption effects (take $\tau_\epsilon =0$):
\begin{equation}
F_l(\eta_A,\vx_A,\epsilon,k) \approx
e^{-\tauc} \bar{f}(\epsilon) \left[\Delta_s j_l(k\chi_\epsilon)+ \frac{k v}{\clh}  j_l''(k \chi_\epsilon)\right].
\label{twoterm}
\end{equation}
The fractional angular power spectrum for one redshift shell is then
\begin{equation}
C_l(z,z) \approx 4\pi e^{-2\tau(z)} \int \ud \ln k \biggl\{ \clp_{\Delta_s}(k,z) [j_l(k\chi(z))]^2 + 2\clp_{v\Delta_s}(k,z) j_l(k\chi(z)) j_l''(k\chi(z)) + \clp_{vv}(k,z)  [j_l''(k\chi(z))]^2\biggr\},
\label{Cl_sharp_exact}
\end{equation}
where ${\cal P}_{\Delta_s}$, ${\cal P}_{v}$, and ${\cal P}_{v\Delta_s}$ are the power spectra of $\Delta_s$, $k v / \clh$, and their cross-correlation\footnote{We define power spectra so that $\la \Delta(x)^2\ra =\int \ud \ln k \clp(k)$.}. For $k\chi$ several oscillations larger than $l$, and smooth power spectra, we can replace the products of the rapidly oscillation Bessel functions with their approximate smooth averages from Appendix~\ref{Bessels}.

One might expect the low-$l$ 21cm fluctuations to be dominated by the
super-horizon scale, post-Newtonian fluctuations at scale
$k=l/\chi$. This is not correct since small-scale fluctuations,
with amplitude growing rapidly with scale as $k^2$,
couple to $l < k \chi$ through the oscillatory tails of the spherical
Bessel functions. The net effect is that, for all $l$ and a single
source plane, the dominant
contribution is from modes that are inside the horizon.
At low $l$, the $C_l$ integral of Eq.~\eqref{Cl} is therefore dominated by much smaller scales, with $k \gg l/\chi$. In this limit the Bessel functions can be approximated with the asymptotic result $j_l(k\chi) \sim \cos\left[k\chi - (l+1)\pi/2\right]/(k\chi)$. Since the power spectrum is quite smooth we can then average over oscillations using $\la |j_l(k\chi)|^2 \ra \sim 1/[2(k\chi)^2]$. We can similarly remove oscillating terms in the second derivative term. Hence on large scales, keeping only the monopole source and redshift distortions and assuming a narrow redshift window, the dimensionless fractional power spectrum is
\begin{equation}
C_l(z,z) \sim 4\pi e^{-2\tau_c} \int \ud \ln k \frac{\clp_s(k,z)}{2k^2\chi(z)^2},
\end{equation}
where $\clp_s(k,z)$ is the power spectrum of $\Delta_s -k v/\clh$ at redshift $z$ (conformal distance $\chi(z)$). Since this is independent of $l$ it corresponds to a white noise spectrum. The intrinsic fluctuations on a scale $k=l/\chi$ are hidden beneath random variations in the large scale distribution of much smaller perturbations. This is a generic feature of the 21cm angular power spectrum that is also true after the dark ages~\cite{Barkana:2004zy}.

Scales inside the horizon at matter-radiation equality have an approximately scale-invariant spectrum (grow logarithmically with $k$). This causes the 21cm power spectrum to flatten out. First consider the case where the window function is sharp,  $\Delta\chi/\chi \ll 1/l$, so the source is from a single redshift. For monopole and velocity sources with power-law spectra, the result for $C_l$ from Eq.~\eqref{Cl_sharp_exact} can be obtained analytically using a result for
integrating products of spherical Bessel functions quoted in Appendix~\ref{Bessels}. In particular, taking the power spectrum to be approximately constant on small scales, we can approximate the dimensionless fractional power spectrum as
\begin{eqnarray}
\frac{l(l+1)}{2\pi}C_l(z,z) &\sim& e^{-2\tau_c} \left[ {\cal P}_{\Delta_s}(\pi l/2\chi(z),z) -\frac{2}{3}  {\cal P}_{v\Delta_s}(3\pi  l/4\chi(z),z) + \frac{1}{5} {\cal P}_{v}(15\pi l/16\chi(z),z)\right].
\label{sharp_approx}
\end{eqnarray}
The numerical factors are consistent with the angular average of Eq.~\eqref{fapprox}. Since the small scale spectrum actually grows logarithmically, the power spectra in Eq.~\eqref{sharp_approx} are approximated by their values at the mean position of the corresponding window function. The Bessel functions are skewed to $k > l/\chi$ so that the mean of a $[j_l(r)]^2/r$ window is at $r=l(l+1)\pi/(2l+1)\sim \pi l/2$. The velocity and cross-power window functions probe even smaller scales $\sim 15\pi l/16$ and $\sim 3\pi l/4$ respectively. Eq.~\eqref{sharp_approx} is accurate at the $10\%$-level from the end of the baryon oscillations to the baryon damping scale at $l\sim 10^6$. The power is overestimated because the underlying power spectra are only growing logarithmically rather than linearly.
 In the (crude) approximation that $\Delta_s \sim \Delta_b \sim \Delta_c$ so that $k v/\clh \sim k v_c/\clh \sim -\Delta_s$, and taking $\clp(k)$ to be constant we have
\begin{equation}
\frac{l(l+1)}{2\pi}C_l(z,z) \sim   \frac{28}{15} e^{-2\tau_c}  \clp_{\Delta_s} (k, z).
\end{equation}
Though not very accurate this result shows the importance of redshift distortions: it is $\sim 28/15 \sim 1.87$ times larger than the equivalent approximate result neglecting them.
%In reality the small-scale spectrum is growing logarithmically with $k$, so~\eqref{sharp_approx} slightly underestimates the small-scale power. When baryon pressure is important the $k\chi>l$ tail gives a small contribution as the power is falling rapidly -- the approximation becomes an overestimate.

On scales where the wavelength is much smaller than the redshift bin width, $l\agt \chi/\Delta\chi$, redshift distortion effects average out and we can instead use the Limber approximation:
\begin{equation}
C_l(z,z) \approx  e^{-2\tau_c} \frac{2\pi^2}{l^3} \int_0^{\chi_*} \chi \ud \chi\, W(\chi)^2
{\cal P}_{\Delta_s}(k=l/\chi(z);\eta_0-\chi(z)),
\end{equation}
where $\chi_*$ is the far end of the window function $W(\chi)$. Since the source power spectrum is nearly scale invariant, $C_l$ therefore scales approximately as $1/l^3$. In other words $l^2 C_l$ is approximately constant with an additional $1/l$ suppression due to line-of-sight averaging through the window. If $W$ is a Gaussian of width $\sigma$, the line-of-sight averaging causes the overall amplitude to scales approximately as $1/\sigma$. The Limber approximation can in fact be used for numerical calculation on scales with $l\gg \chi/\Delta\chi$ where it becomes accurate.

One very small scales where modes are inside the baryon sound horizon at recombination, $k c_s(\eta_*)\eta_*\agt 1$,  the baryon pressure becomes important and $\Delta_H$ differs significantly from $\Delta_c$ even at late times. We discuss approximate analytic solutions for the evolution of $\Delta_b$ in Appendix~\ref{baryons}.
In the small scale limit where $k^2c_s^2 \gg \clh^2$ the pressure and gravitational forces approximately balance, and $\Delta_b/\Delta_c \sim \clh^2/(k^2 c_s^2)$. Since $\Delta_c$ is roughly scale invariant, and neglecting the effect of the baryon pressure on the CDM evolution, this implies $l^2 C_l \propto 1/l^4$, giving a characteristic sharp fall-off in power on very small scales (there is an additional power of $1/l$ for window-function line-of-sight averaging). Note the observations on such small scales are unlikely to be possible  at high redshift over most of the sky due to scattering by turbulent galactic and solar-system plasma~\cite{Rickett90}. The non-zero linewidth also becomes important one these scales, so our results in the approximation of a monochromatic source will overestimate the power.

\section{Numerical results}
\label{sec:numerics}

\begin{figure}
\begin{center}
\epsfig{figure=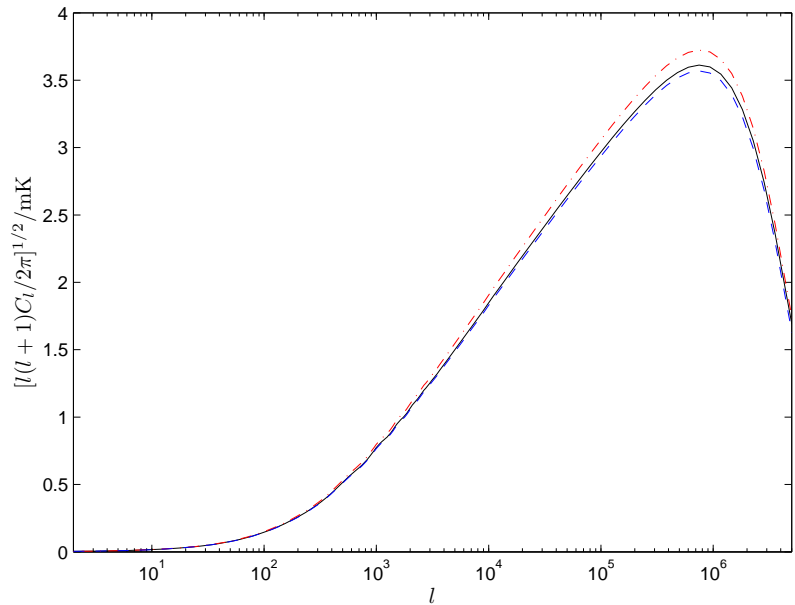,width=10.5cm}
\caption{The effect of perturbations to the optical depth and ionization fraction on the 21cm power spectrum at $z=50$ with a sharp window function. The solid line shows our main result, the dashed-dotted line is the larger result using the a zeroth-order expansion in $\tau_\epsilon$, the dashed line is the lower result if the effect of ionization fraction perturbations on the gas temperature evolution is neglected. The fractional change in the spectrum is roughly the same on all scales where baryon pressure is negligible.
\label{termssmall}}
\end{center}
\end{figure}

\begin{figure}
\begin{center}
\epsfig{figure=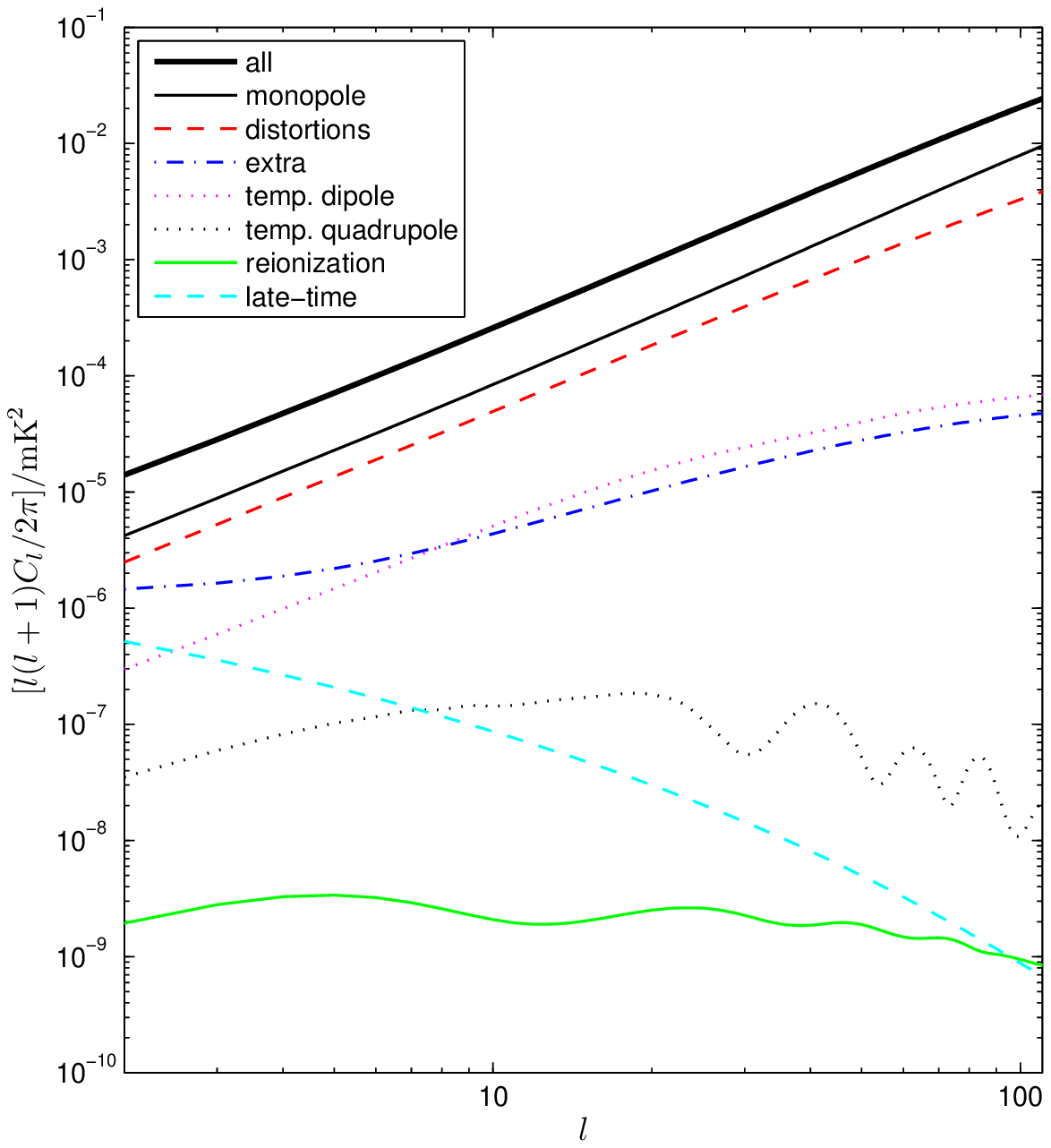,width=8.5cm}
\epsfig{figure=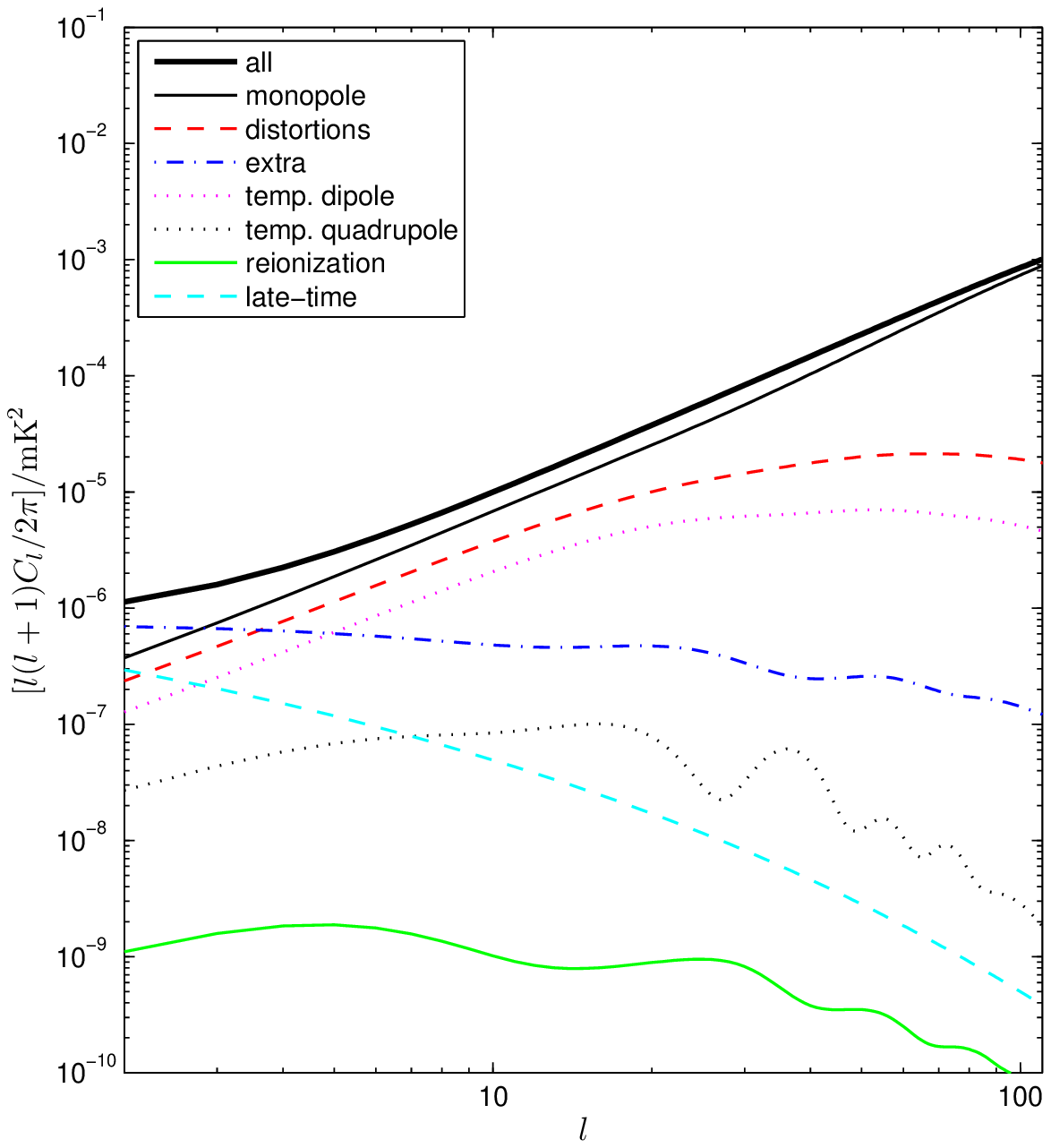,width=8.5cm}

\caption{Auto power spectra for the various terms in the large-scale 21cm power spectrum. Left: narrow window function at $z=50$ with $\Delta\nu = 0.01\Mhz$ ($T_b = -26\mK$); Right: broad window function at $z=40$ with $\Delta\nu = 5\Mhz$ ($T_b = -15\mK$, $\sigma_z \sim 6$). Terms are calculated in the synchronous gauge, and `extra' includes all effects not included in other curves; standard redshift distortions are defined here by the second term in Eq.~\eqref{twoterm}. The late-time curve is the ISW contribution from line-of-sight redshifting. The reionization curve is the result from sources at reionization, the other curves include the main $e^{-2\tau_c}$ damping effect. The standard calculation includes only monopole and redshift distortion terms; the difference from the full result is $\sim 1\%$ at $l\alt 50$ growing to a few percent at low l for the narrow window function. For the broad window function that averages down small-scale power extra terms change the total by $\agt 1\%$ at $l<100$ (growing to $\sim 40\%$ at low $l$).
Note that the total spectrum is not just the sum of the other autocorrelation terms since it also includes all cross-terms. }
\label{terms}
\end{center}
\end{figure}

\begin{figure}
\begin{center}
\epsfig{figure=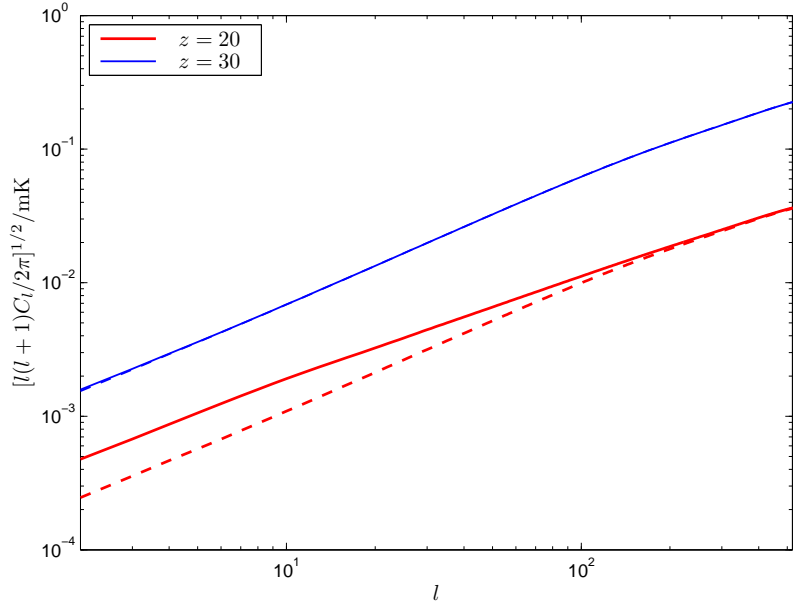,width=10.5cm}
\caption{The large-scale 21cm power spectrum at $z=20$ and $z=30$ ($\Delta\nu = 0.01\Mhz$) if there were no Lyman-$\alpha$ sources, shock heating, minihaloes, or other non-linear effects. Solid lines are the full linear result, dashed lines include only monopole and redshift-distortion sources. The difference is dominated by the baryon-photon velocity term.}
\label{lowz}
\end{center}
\end{figure}

\begin{figure}
\begin{center}
\epsfig{figure=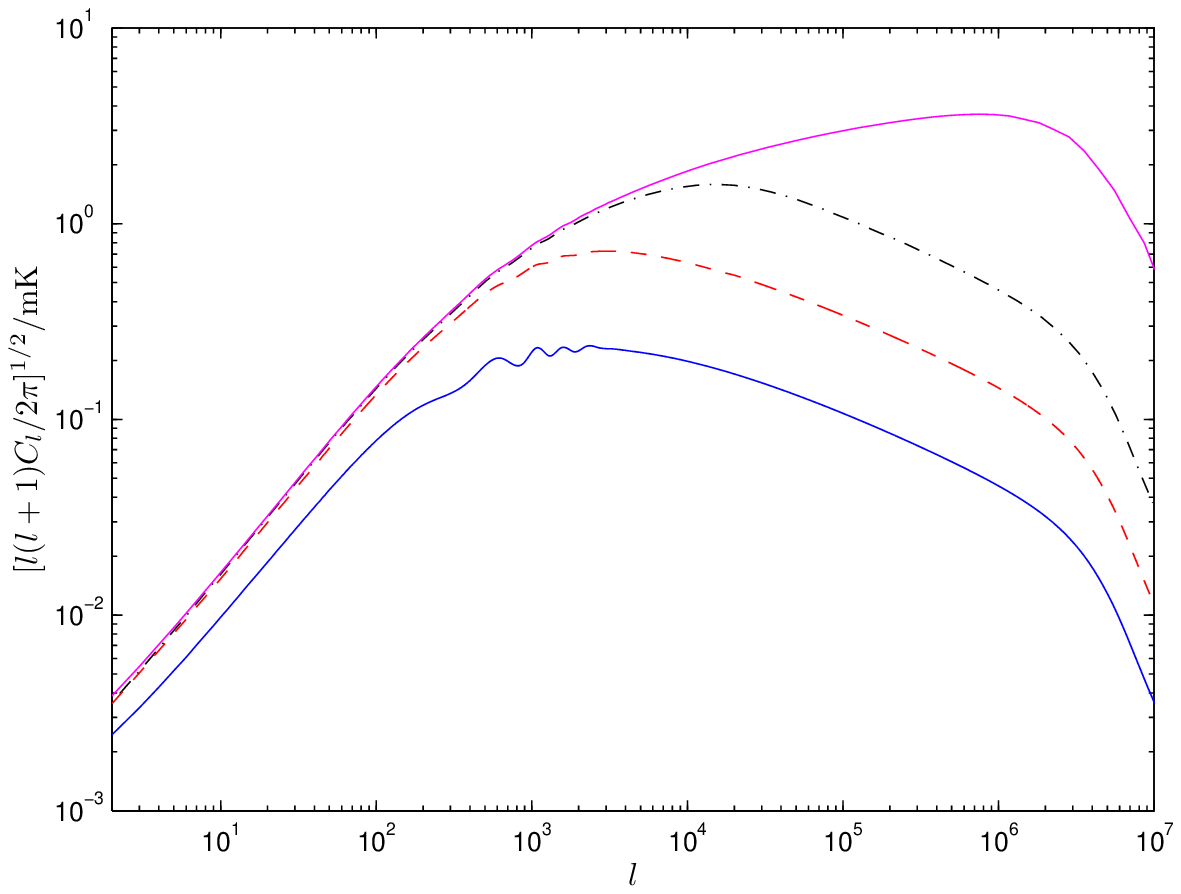,width=10.5cm}
\caption{The 21cm power spectrum at $z=50$ for $\Delta\nu = \{1, 0.1, 0.01, 0\}\Mhz$ (bottom to top). Large widths suppress the redshift-distortion contribution and allow the baryon oscillations to be seen. All show characteristic damping due to line-of-sight averaging over the bin width at $l\agt \chi/\Delta\chi$, and the effect of baryon pressure at $l\agt 5\times 10^{6}$.
\label{binwidths}}
\end{center}
\end{figure}

For our numerical results we assume a standard concordance flat adiabatic CDM ($\Omega_c h^2=0.104$) model with a constant primordial spectral index ($n_s=0.95$) and optical depth to Thomson scattering $\tau=0.09$. We take baryon density $\Omega_b h^2=0.022$, Hubble parameter $73 \text{km} \text{s}^{-1} \text{Mpc}^{-1}$ and initial curvature perturbation power on $0.05 \Mpc^{-1}$ scales $A_s=2.04\times 10^{-9}$. We neglect the neutrino masses, which do not have a large effect for source planes at high enough redshift that the neutrinos are still relativistic. We take our window function to be a Gaussian of width $\Delta\nu$ in frequency over the observed brightness.

Fig.~\ref{termssmall} shows the effect on the angular power spectrum of allowing for self-absorption and ionization fraction perturbations. The finite non-zero optical depth lowers the amplitude both because the background signal is lower and because the optical depth from an overdensity is higher than from the background; each effect is about $\tau_\epsilon/2\sim 1.5\%$ in amplitude, giving an overall suppression of $\sim 6\%$ in power. Ionization fraction perturbations increase the power by about $2\%$ on all scales: overdensities recombine more fully and hence have less Compton-coupling to the CMB and hence lower spin temperature. These effects are important on all scales and should be included in any accurate calculation, though note that they are comparable to others effects that we have neglected because of the simple velocity-independent spin-temperature approximation (see Ref.~\cite{Hirata:2006bn}).

On small scales the post-Newtonian and extra velocity terms can be neglected to good accuracy. On large scales they can be more important. In Fig.~\ref{terms} we show the contribution of the auto-variance of various terms to the total large-scale power spectrum at a given redshift. As expected the dominant contributions are still from the monopole source fluctuations and redshift distortions\footnote{Note that at the percent level it is important to use the baryon rather than CDM velocity when calculating the redshift distortions.}. Except on very large scales the next most important term is from the CMB dipole in the rest frame of the gas, which gives percent level contribution at $l<50$ for the redshift shown here, but is negligible on much smaller scales\footnote{Note that although the CMB dipole signal has only a small effect on the dark-age 21cm power spectrum, it may make a larger contribution to the correlation with other sources, for example the cross-correlation with the CMB temperature during reionization (c.f. Ref.~\cite{Alvarez:2005sa}).}. At ($l<100$) the contribution from the potential and other velocity terms are also not entirely negligible. The contributions from the CMB temperature anisotropy above the dipole, and reionization re-scattering sources are completely negligible on all scales. At lower redshifts the background signal becomes smaller, and the relative importance of the terms changes. The background signal depends on $\bar{T}_s-\bar{T}_\gamma$, but some of the perturbation sources depend only on the 21cm optical depth and are non-zero even when the spin temperature is equal to the CMB temperature. As an extreme example, Fig.~\ref{lowz} shows the relatively large contribution from the photon-baryon dipole at $z=20$ on large scales if there were no additional sources from non-linear structures.

Note that just because something does not show up in a narrow frequency window auto-power spectrum at a given redshift does not mean that it is necessarily negligible. The correlation between source planes at a given $l$ falls off very rapidly once the plane separation is greater than characteristic perturbation size $\chi/l$. Extra information may therefore be available in the cross-power spectra, particularly about small large-scale signals that are correlated between redshift bins. The effect of different frequency window functions is shown in Fig.~\ref{binwidths}. Here the baryon oscillations only show up when the window is wide enough to damp down the large smaller-scale fluctuations so that the power on baryon oscillation scales is not dominated by contributions from smaller scales. When narrow frequency windows are used this information is hidden
in the cross-correlation structure of the different source planes.

The white-noise signal on large scales can be reduced by averaging over many redshift slices. Figure~\ref{terms} shows the relative importance of the various terms on large scales when a broad redshift window function is used. Redshift distortions from scales smaller than the bin width are suppressed, and the large-scale white-noise monopole signal is reduced because of the line-of-sight averaging of small scale power. The relative importance of the additional terms is therefore larger. This raises the question of whether the large-scale 21cm signal can be useful, for example to learn about large scale power or as a source for the integrated Sachs-Wolfe effect\footnote{We thank Jeff Peterson for raising this question during a talk.} (ISW). There are two main problems. Firstly, a very broad window function is required to make the extra terms comparable to the monopole source, and residual monopole and redshift distortion signals will generally dominate. Secondly, since the dark age redshift shells are $\agt 3/4$ of the comoving distance to the last scattering surface, large-angle correlations, such as those due to Sachs-Wolfe potential redshifting, will be strongly correlated between redshift slices and correlated with the large-scale CMB. As an averaged source plane for the ISW the 21cm signal therefore has at least as much large-scale `noise' from other sources as the CMB and hence offers little extra information.

Decorrelation with source plane separation is a powerful way to try and separate intrinsic and foreground signals due to the much smoother signal (as a function of frequency) expected from most foregrounds~\cite{Zaldarriaga:2003du,Datta:2006vh}. Detection of small non-foreground cross-correlations is therefore particularly challenging.

%Cut this figure
%Fig.~\ref{cross} shows the significant change to the large-scale redshift-cross-correlation %signal due to additional sources.

\section{Polarization}
\label{sec:pol}

\begin{figure}
\begin{center}
\epsfig{figure=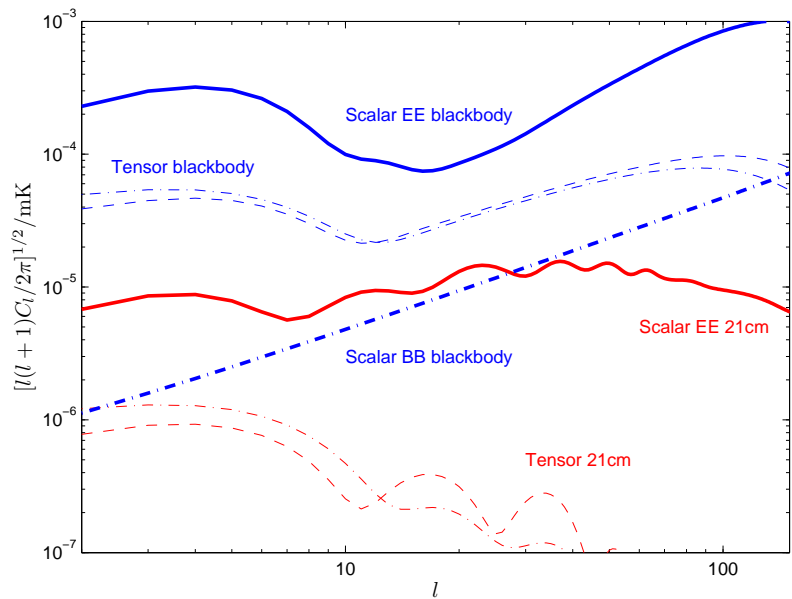,width=10.5cm}
\caption{The 21cm polarization power spectra from $z=50$ with $\Delta\nu = 0.01\Mhz$. Thick lines show the adiabatic scalar-mode signal, thin lines from $r\equiv A_T/A_s=0.1$ scale-invariant tensor modes. For comparison, the instrinsic CMB
polarization signals (labelled blackbody) are also plotted: the scalar $EE$
blackbody curve is the lensed E-mode polarization from scalar modes
and the scalar $BB$ blackbody curve is the lens-induced B-mode power; the
tensor blackbody curves are the CMB polarization power spectra from gravity
waves. For both intrinsic CMB and 21cm fluctuations, E-mode polarization is
shown with solid or dashed lines and B-modes with dash-dotted lines.
There are no 21cm scalar B-modes within our approximations.
Reionization is assumed to be fairly sharp with optical depth $\tau=0.09$ ($z_{\text{re}}\sim 12$).
%For comparison, the curves that end at the top right are the intrinsic CMB temperature signal (including lensing), the lower curves show the 21cm signal. Dashed is E-polarization, dash-dotted B-polarization.
\label{pol}}
\end{center}
\end{figure}

A quadrupole anisotropy in the 21cm signal at reionization can generate polarization by Thomson scattering~\cite{Babich:2005sb}. The signal is expected to be very small, but perhaps worth considering as polarization may be very useful to help with foreground cleaning of scalar modes. In principle the tensor mode signal can also be used as a cross-check on CMB temperature and polarization constraints on models of inflation.

Following standard methods~\cite{Hu:1997hp,Challinor:2000as} it is straightforward to modify our scalar equations to include polarization in order to calculate the E-mode (gradient-like) spectrum generated at reionization. For our idealized analysis we neglect any effects due to magnetic fields, anisotropy of the hydrogen triplet state distribution and inhomogeneity of reionization. Typical numerical results are shown in  Fig.~\ref{pol}.

Gravitational waves (tensor modes) are known to be subdominant to the scalar modes, but can also source anisotropies by their anisotropic shearing. There are two mechanisms. Firstly, the metric shear can directly change the 21cm photon frequency along the line of sight, which distorts the emission shell in much the same was as the redshift distortions and line-of-sight effects do for the scale modes. Secondly, the CMB temperature at absorption will be anisotropic due to gravitational waves between the absorber and the last scattering surface: this causes an anisotropy in the 21cm absorption. At reionization the quadrupole component of these anisotropies source both E and B polarization. The signal is a small fraction of the blackbody tensor signal because the optical depth for 21cm emission is only $~\sim 0.02$.  Note that since there is no intrinsic 21cm polarization before reionization the lensing-induced 21cm B modes are much smaller.

\section{Non-linear evolution}
\label{sec:nonlin}

\begin{figure}
\begin{center}
\epsfig{figure=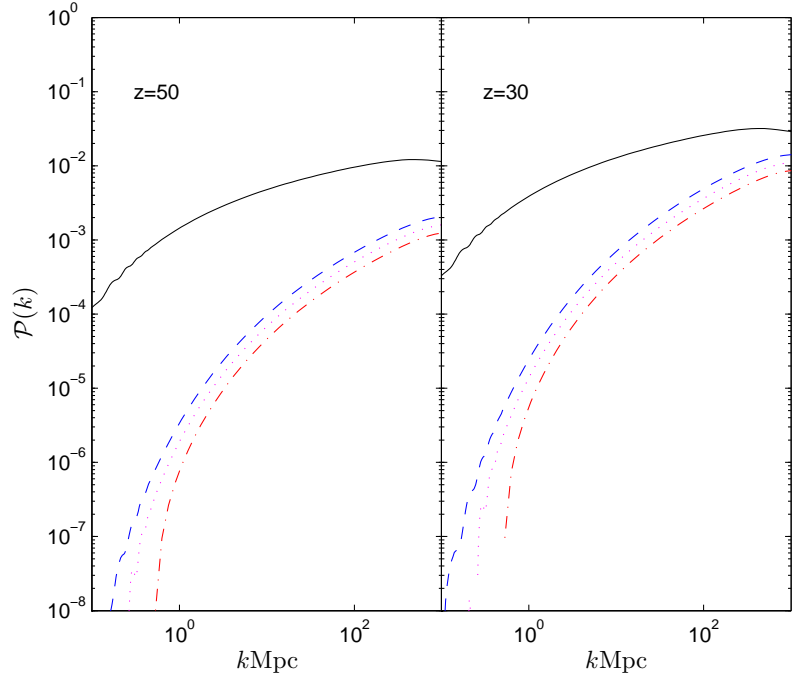,width=10.5cm}
\caption{Power spectrum of the dark matter $\clp(k)$, linear result (solid) and approximate second-order terms from Appendix~\ref{nonlinear}: $\clp^{\delta\delta}(k)$ (dashed), $\clp^{v v}(k)$ (dot-dashed), $-\clp^{v\delta}(k)$ (dotted). On large scales the second-order contributions become negative (not shown).
\label{nonlinpk}}
\end{center}
\end{figure}

\begin{figure}
\begin{center}
\epsfig{figure=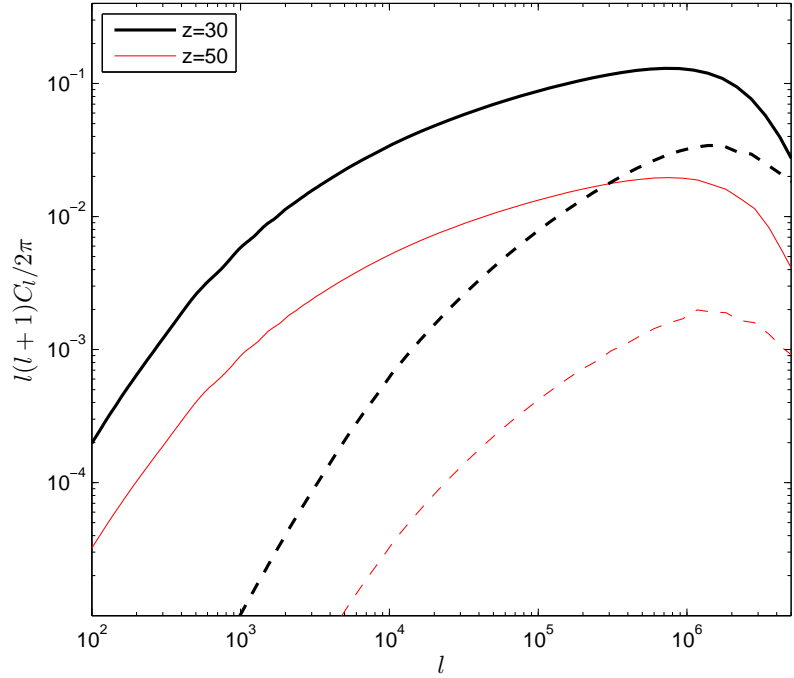,width=10.5cm}
\caption{The linear 21cm angular power spectrum at $z=30$ and $z=50$ (solid lines), and the approximate higher-order correction from non-linear evolution (dashed lines). Window functions are taken to be sharp ($\Delta\nu = 0$).
\label{nonlincl}}
\end{center}
\end{figure}

Although the dark age perturbations are quite small, non-linear effects can still be important. This is clear from Fig.~\ref{transfer}, where the perturbation amplitude on Jeans' scales at $z=50$ corresponds to density perturbations of order ten percent. We shall estimate the effect of non-linear evolution using Newtonian perturbation theory in the approximation that vorticity and decaying modes are unimportant~\cite{Vishniac83,Makino92,Jain:1993jh,Scoccimarro:2004tg,McDonald:2006hf}. The result should be quite accurate for the dark matter density as perturbation theory gives good results in the mildly non-linear regime. Assuming an initially Gaussian field one might expect the non-linear contribution to the power spectrum to be of order $|\Delta_c|^4$, corresponding to a correction of about one percent. However, on very small scales, there are many larger scale $k$-modes that contribute
to the local density, giving a total effect from all mode couplings that
is significantly larger than the simple estimate.

For our approximate analysis we focus on the CDM perturbations during matter domination, neglecting any effect due to the baryons.
The first non-linear contribution to the power spectrum comes from two terms, $\clp_{22}$ coming from the square of the second-order perturbation, and $\clp_{13}$ coming from the cross-term between the first and third order perturbations. We give the results for the CDM density, velocity and cross-correlation power spectra in Appendix~\ref{nonlinear}; See Fig~\ref{nonlinpk} for typical numerical results.

To calculate the effect of non-linear evolution on the 21cm power spectrum one should really evolve the full coupled baryon,  temperature and CDM equations to third order. This is beyond the scope of this paper. Here we instead make the approximation that the 21cm monopole source and baryon velocity power spectra have the same fractional contribution from non-linear evolution as do the CDM power spectra. Using Eq.~\eqref{Cl_sharp_exact} we then get the results shown in Fig.~\ref{nonlincl}. Nonlinear effects are important at the few-percent level on small scales even at redshift $z\sim 50$. At redshift $z\sim 30$ there is a $\sim 10\%$ correction to the power spectrum at $l \sim 10^5$. A more accurate analysis at redshift $z\sim 30$ may need to account for gas shock heating, minihaloes, or even rare first sources (see e.g. Ref.~\cite{Furlanetto:2006jb} and references therein).

\section{Lensing}

\label{sec:lens}

\begin{figure}
\begin{center}
\epsfig{figure=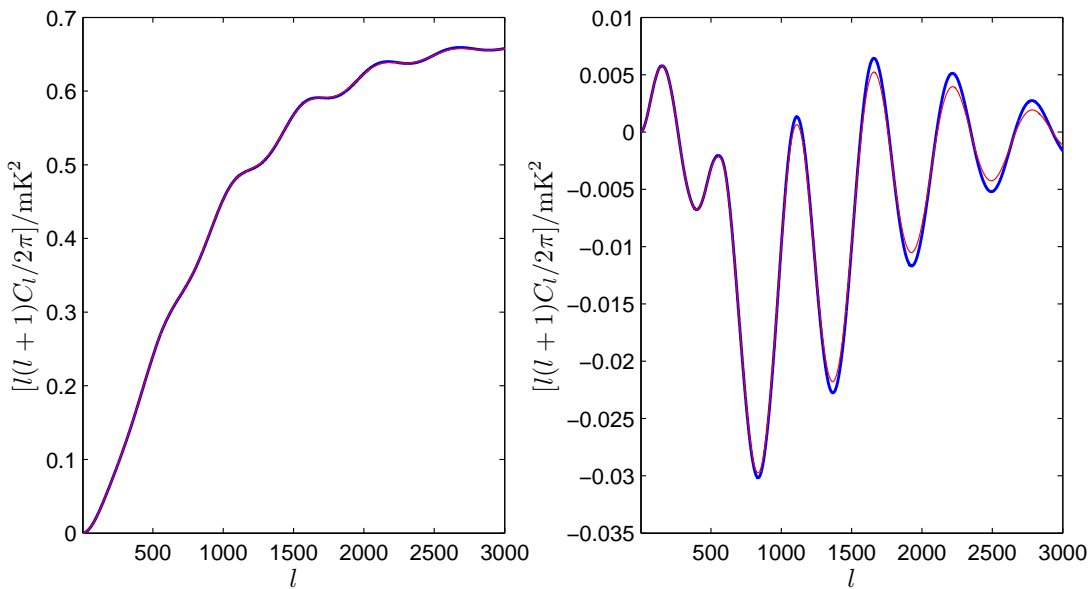,width=14.5cm}
\caption{The 21cm power spectrum for bins at $z=50$ and $z=52$ ($\Delta\nu = 0.1\Mhz$), with lensing (thin lines) and without lensing (thick lines). The left figure shows $C_l(z,z)$, the right figure shows the small but non-negligible cross-correlation.
\label{lensing}
}
\end{center}
\end{figure}

The line emission spectrum is not affected by lensing at first order. However perturbations along the line of sight can produce a non-negligible higher-order effect. The effect on the two- and three-dimensional power spectra has been analysed in detail in Ref.~\cite{Mandel:2005xh}.
Lensing acts rather like a convolution of the angular power spectrum with the deflection-angle power spectrum. Since the 21cm power spectrum is rather smooth, the effect is significantly smaller than on the CMB. Nonetheless the baryon wiggles can be smoothed at well above the cosmic variance level. Here we note a simple, non-perturbative approach to calculating the effect on the angular power spectrum. This is based on the lensed correlation function method often applied to the CMB temperature and polarization~\cite{Seljak:1996ve,Challinor:2005jy,Lewis:2006fu}.

Since the lensing field is nearly linear the effect is well described in terms of the lensing deflection angle power spectrum $C_l^\alpha(z,z')$. In linear theory this is given in terms of the primordial power spectrum  $\clp_{\mathcal{R}}(k)$ and transfer functions $T_\Psi(k;\eta)$ for the Weyl potential,
$\Psi \equiv (\phi + \psi)/2$,
by (see e.g. Ref.~\cite{Lewis:2006fu})
\begin{multline}
C_l^\alpha(z,z')  = 16\pi l(l+1)\times \\
\int \frac{\ud k}{k}\, \clp_{\mathcal{R}}(k)
%\\\times
\left[\int_0^{\chi(z)}\!\!\ud \chi\,
T_\Psi(k;\eta_0-\chi) j_l(k\chi) \left(\frac{\chi(z)-\chi}{\chi(z)\chi}\right)\right]
\left[\int_0^{\chi(z')} \!\!\ud \chi'\,
T_\Psi(k;\eta_0-\chi') j_l(k\chi') \left(\frac{\chi(z')-\chi'}{\chi(z')\chi'}\right)\right],
\label{cpsi_transfer}
\end{multline}
where $\chi(z)$ is the conformal distance to the window at redshift $z$. We assume the window function is narrow compared to the distance so that the lensed sources can be approximated as a single source plane. For nearby bins the terms in square brackets are approximately equal.

In terms of $C_l^\alpha(z,z')$ and the unlensed spectrum $C_l(z,z')$, the lensed power spectrum is given to very good accuracy by~\cite{Lewis:2006fu}
\begin{equation}
\tilde{C}_{l'}(z,z') \approx  \sum_l \frac{2l+1}{2} C_l(z,z')
\int_{-1}^1 \ud \cos\beta \,
d^{l'}_{00}(\beta)
e^{-l(l+1)\sigma^2(\beta)/2}
%\\\times
\sum_{n=-l}^l  I_n\left[l(l+1) \Cgltwo(\beta)/2\right] d^l_{n-n}(\beta) ,
\label{curvt}
\end{equation}
where $I_n$ is a modified Bessel function, $d^l_{mn}$ are reduced Wigner functions, $\sigma^2(\beta) \equiv \Cgl(0) - \Cgl(\beta)$, and
\begin{eqnarray}
 \Cgltwo(\beta) \equiv
 \sum_l \frac{2l+1}{4\pi}C_l^\alpha(z,z') d^l_{-1 1}(\beta)  , \nonumber \\
\Cgl(\beta) \equiv
\sum_l \frac{2l+1}{4\pi} C_l^\alpha(z,z') d^l_{1 1}(\beta).
\label{cgl_def_curv}
\end{eqnarray}
The sum over $n$ is dominated by the lowest terms with $|n|\approx 0$. For a $\Delta\nu = 1 \Mhz$ window at $z=50$, the lensing-induced smoothing of the baryon oscillations can be nearly a percent, but elsewhere the effect on the angular power spectrum is generally very small because it is smooth on the scale of the width of the deflection angle power spectrum ($\Delta l \sim 100$). Figure~\ref{lensing} shows the significant smoothing effect on the cross-correlation power spectrum on the scale of the baryon oscillations.

In addition to the small effect on the power spectrum calculated here and in
Ref.~\cite{Mandel:2005xh}, lensing also makes the distribution non-Gaussian. Combined with tomographic information this can be used for lensing reconstruction~\cite{Zahn:2005ap,Metcalf:2006ji,Hilbert:2007jd}.

\section{Other non-linear effects}
\label{sec:othernonlin}
On small scales much larger than the line width the non-linear angular distribution can be obtained from integrating
Eq.~\eqref{eq:exactBoltzmann}. In the Newtonian approximation, neglecting self-absorption and lensing, and dropping small $\vnhat\cdot\vv$ terms that are unimportant on small scales, we have the approximate non-linear result
\begin{equation}
f(\vnhat,z) \approx  e^{-\tau_c} \sum_{\chi'} \left.\frac{(1+\Delta_s)\bar{\rho}_s}{1+(\vnhat\cdot \partial \vv/\partial\chi)/\clh}\right|_{\chi'},
\end{equation}
where $\chi'$ are displaced positions satisfying $\chi' = \chi(z) - \vnhat\cdot \vv(\chi')/\clh$. In the linear approximation we used $\Delta_s(\chi') = \Delta_s(\chi)$, however this will not be a good approximation if the Doppler displacement is comparable to the scale of the perturbation. The RMS velocity is around $3\times  10^{-4}$ at redshift 50, corresponding to a displacement of $\sim 0.3 \Mpc$, and hence suggesting that non-linear Doppler displacement effects may become large at $k\sim 10 \Mpc^{-1}$. However the small-scale \emph{power spectrum} will be unchanged under small bulk radial displacements: the spectrum at wavenumber $k$ is only really sensitive to the \emph{difference} in the radial displacement over a distance comparable to $1/k$. Since the velocity power spectrum falls with scale the effect is much smaller than indicated by the above estimate. The displacement on scale $k$ is given by $\delta\chi \sim \vnhat\cdot\vv/\clh \sim \Delta_c/k$, and thus $k\delta\chi$ will only become large on scales where $\Delta_c$ does.
%AL: not sure about this
%Unlike the case of non-linear evolution discussed above, where the local density and hence non-linear evolution depends on all the larger scale modes, there is no reason to expect the velocity effect to have much coupling to larger scales.
Coupling to smaller scales will give an effective velocity dispersion similar to a scale-dependent thermal line broadening, and will tend to reduce the power. Since the velocity power falls with $k$ this is never a large effect. The total non-linear effect is therefore likely to be comparable to the estimated contribution from just the non-linear evolution.

A perturbative analysis using third-order Newtonian perturbation theory is given in Ref.~\cite{Heavens:1998es}, but a detailed investigation of the effect on the small scale 21cm power spectrum is beyond the scope of this paper. Since higher order corrections to $\Delta_s(\chi')$ are large on small scales, but the correction to the power spectrum is small due to the large coherence length of the displacement, a perturbative treatment will involve delicate cancellations between different independently large terms.

%
%The RMS velocity difference at $\alt 0.3 \Mpc$-scales is $\sim 5\times 10^{-6}$, suggesting the effect may be well below the %linear result on all scales.
%Nonetheless the corrections could be quantitatively important. A detailed investigation is beyond the scope of this paper and may be covered in a future publication. There are also numerous other non-linear effects, though most are expected to be small.

\section{Conclusions}
\label{sec:conc}

The 21cm signal from the dark ages is potentially a powerful probe of cosmology. We have derived full linear-theory results for the angular power spectra, and confirmed that, for most purposes, standard approximations including only monopole sources and redshift distortions are accurate at the percent level. However additional velocity and potential terms can be non-negligible on very large scales, and optical depth and ionization fraction perturbations have a significant percent-level effect on all scales. Non-linear evolution can be a many-percent effect on small scales even at high redshift, and gravitational lensing can also be important, though this is easily modelled.

 In this paper we have focussed on theoretical issues. The observational problems are formidable, requiring many square-kilometers of collection area, low radio-noise, and, for high redshifts, getting outside the atmosphere to avoid the opaque ionosphere; see Refs.~\cite{Zaldarriaga:2003du,Furlanetto:2006jb,Carilli:2007eb,Peterson:2006bx} for further discussion. The additional hurdles to measure the distinctive polarization signal that we calculated, which is below the intrinsic blackbody signal, are quite possibly insurmountable.

The 21cm spectrum is much more sensitive to the residual ionization fraction after recombination than the CMB temperature: the ionization fraction governs the evolution of the matter temperature which in turn affects the background brightness and the evolution of the perturbations. Current modelling of recombination is probably insufficiently detailed to be able to compute the 21cm power spectrum to an accuracy of more than a couple of percent as there are a wealth of subtle effects that need to be modelled to calculate hydrogen recombination accurately~\cite{Switzer:2007sq}. As an example of current uncertainties, using the result of Ref.~\cite{Rubino-Martin:2006ug} rather than RECFAST~\cite{Seager:1999km} for the post-recombination ionization fraction changes the 21cm brightness temperature at the two-percent level. If future observations are to achieve the $\sim O(10^{-5})$ accuracy on cosmological parameters that is possible in principle, a very detailed analysis will be required of the full electron, atomic state and  velocity distributions, with all interactions, extending the work of Ref.~\cite{Hirata:2006bn}. Furthermore, even at redshift of $50$ the Jeans'-scale density perturbations are already $\clo(0.1)$ and non-linear corrections are important at the many percent level. We estimated this effect with an approximate second-order CDM perturbation theory calculation, however accurate results will require a detailed study of the full baryon-CDM evolution at second and third order or beyond.

Although we have concentrated on the power spectrum from the dark ages, much of our formalism can be adapted straightforwardly to the reionization epoch given models for the background evolution and additional sources. Our numerical code is publicly available\footnote{\url{http://camb.info/souces}}.

\section{Acknowledgements}
AL acknowledges a PPARC Advanced fellowship and AC a Royal Society University Research Fellowship. AL thanks Oliver Zahn, Steven Gratton, Ue-Li Pen and Jeff Peterson for discussion.

\appendix

\section{Baryon perturbation evolution}
\label{baryons}
After recombination, Compton scattering has no significant effect on the baryon density evolution and the fractional synchronous gauge baryon perturbation $\Delta_b$ obeys the equation
\begin{eqnarray}
\frac{\partial^2 \Delta_b}{\partial\eta^2} + \clh\frac{\partial \Delta_b}{\partial\eta} + k^2 c_s^2 \Delta_b &=&  4\pi Ga^2 \sum_i (\delta\rho_i + 3\delta p_i)\nonumber\\
&\approx & \frac{3}{2}\clh^2 \Delta_m,
\label{baryon_evolve}
\end{eqnarray}
where $(\rho_b+\rho_c) \Delta_m = \rho_c\Delta_c +\rho_b\Delta_b$ and the second line assumes matter domination.
We make the approximation that $\rho_c \gg \rho_b$ so the baryons have no effect on the CDM evolution and we can use the usual result in matter domination that $\Delta_m(\eta)=\Delta_c(\eta) = \Delta_c(\eta_*)\eta^2/\eta_*^2$. On small scales before recombination the baryons are tightly coupled to the photons and silk damping erases perturbations so that at recombination ($\eta\sim \eta_*$) we have $\Delta_b(\eta_*) \sim 0$ and $\dot{\Delta}_b(\eta_*)\sim 0$. On scales where baryon pressure is irrelevant, $k c_s/\clh \ll 1$ at all times, the solution after recombination is
\begin{equation}
\Delta_b/\Delta_c = 1 - 3\left(\frac{\eta_*}{\eta}\right)^2 + 2\left(\frac{\eta_*}{\eta}\right)^3.
\label{baryon0}
\end{equation}
On smaller scale the baryon pressure is important via the $k^2 c_s^2$ term in Eq.~\eqref{baryon_evolve}.
The sound speed decays approximately as $1/\eta$ while Compton scattering couples the gas temperature to the CMB temperature. Once coupling becomes ineffective at $\eta\sim \etaad$ the gas cools adiabatically and $c_s$ decays as $1/\eta^2$. We can solve analytically for the evolution of the baryon perturbation for both limiting behaviours of the sound speed. For $\eta_* < \eta < \etaad$ the result is
\begin{equation}
\Delta_b/\Delta_c =\frac{6}{x^2+6} \left[1 - \left(\frac{\eta_*}{\eta}\right)^{5/2}\left(
\cos[\sqrt{x^2-1/4}\ln(\eta/\eta_*)] + \frac{5}{2\sqrt{x^2-1/4}}
\sin[\sqrt{x^2-1/4}\ln(\eta/\eta_*)]\right)\right],
\label{baryon1}
\end{equation}
where we defined $x\equiv k c_s(\eta_*) \eta_*$ and assumed $x>1/2$.
The analytic continuation of this result to $x=0$ agrees with
Eq.~\eqref{baryon0}. A similar result is given in Ref.~\cite{Nusser:2000zz}, which also discusses the generalization to non-negligible baryon fraction.
On large scales where modes are outside the baryon sound horizon at recombination ($x \ll 1$) the perturbations fall into the CDM potential wells in the order of a Hubble time and $\Delta_b \rightarrow \Delta_c$.  On small scales the baryon pressure causes the perturbations to oscillate once they reach pressure support. The relative amplitude of the oscillations about the midpoint decays as $(\eta_*/\eta)^{5/2}$, so by the time of adiabatic cooling we may neglect this term to $\sim 10\%$ accuracy. The adiabatic cooling equation is a forced harmonic oscillator equation in $1/\eta$ with solution involving sine and cosine integrals that can be written (for $\eta > \etaad$) as~\cite{Abramowitz}
\begin{equation}
\Delta_b = A\cos(u) + B\sin(u) + \Delta_c\left[ 1 - u^2\int_0^\infty \ud t \frac{t e^{-u t}}{t^2+1}\right],
\end{equation}
where $u\equiv x \etaad/\eta = k c_s(\eta)\eta$ and constants $A$ and $B$ are defined by the initial conditions. The term in square brackets monotonically decreases from one as $k$ (and hence $u$) increases, and describes the main effect of the pressure.
Smoothly matching to Eq.~\eqref{baryon1} (with the oscillation dropped)
assuming a sharp transition at $\etaad$, the result can be written as
\begin{equation}
\Delta_b/\Delta_c = 1 - u^2 \int_0^{x-u} \frac{\cos t}{t+u} dt -
\frac{u^2}{6+x^2} \left(\cos(u-x) + \frac{4+x^2}{x}\sin(u-x)\right).
\label{baryon2}
\end{equation}
As a function of $k$ this solution reproduces the qualitative fall-off in power show in Fig.~\ref{transfer}, though the decaying nature of the solution is well hidden in Eq.~\eqref{baryon2}.
The small-scale asymptotic form for $k \rightarrow \infty$ ($u,x \gg 1$) is
\begin{eqnarray}
\Delta_b/\Delta_c &=&\frac{6}{u^2} - 12\frac{u^2}{x^5} \sin(x-u) + \dots\\
&=& \frac{6}{[k c_s(\eta)\eta]^2}\left[1 - \clo\left(\frac{1}{x}\frac{\eta_a^4}{\eta^4}\right) \right].
\label{baryon3}
\end{eqnarray}
As expected the transfer function falls off as $\sim \clh^2/(k^2 c_s^2)|_\epsilon$ on small scales with suppressed oscillations. This power-law fall off will only hold on scales where baryon diffusion can be neglected; on ultra-small scales the spectrum will be exponentially damped. Baryon pressure becomes important at $k_{\text{pressure}}\sim 1/(c_s(\eta_*)\eta_*)$, diffusion damping will become important at $k_{\text{diffusion}} \sim k_{\text{pressure}} (\eta/\eta{_\text{coll}})_\text{min}^{1/2}$ where $\eta{_\text{coll}}$ is the atomic collision time, which is always much smaller than $\eta$ (see Fig.~\ref{backevolve}).

At redshifts of $30$ to $100$ residual Compton coupling actually gives $c_s$ scaling as $\eta^{-1.9}$ to $\eta^{-1.6}$, and approximating the transition in cooling behaviour as sharp may be expected
to be a poor approximation on scales such that $k c_s(\etaad) \etaad \gg 1$. In this
limit, the following approximation can be used:
\begin{equation}
\Delta_b(\eta)/\Delta_c(\eta) = \frac{6}{\eta^3} \int_{\eta_*}^{\eta}
\frac{\eta'}{k\sqrt{c_s(\eta)c_s(\eta')}} \sin\left(\int_{\eta'}^{\eta}
k c_s(\eta'') \, \ud\eta'' \right) \, \ud \eta' .
\end{equation}
This is derived from the WKB solutions of the homogeneous equation with
Greens method. It is an exact solution for all $k$ when the gas is adiabatically cooling,
but only holds for $k c_s(\eta_*)\eta_* \gg 1$ when Compton heating is
still important.
It can be shown that this result agrees with
Eqs.~(\ref{baryon1}) and~(\ref{baryon2}) in the limit $x \gg 1$ (and
for a sharp transition to adiabatic cooling).

\section{General gauge and numerical calculation}
\label{numerics}
We perform numerical calculations in the synchronous gauge. For convenience we give the equations in a general gauge here. The multipole equations in the 1+3 conventions of Refs.~\cite{Challinor:1998xk,Lewis:2002nc} are
\begin{multline}
\dot{F}_l + \frac{k}{2l+1}  \left[(l+1) F_{l+1}
- l F_{l-1}\right]
+\left[\delta_{2l}\numfrac{2}{15} k\sigma - \delta_{0l}\dot{h} -\delta_{1l} \numfrac{1}{3}kA\right] \epsilon\partial_\epsilon \bar{f} \\
= a\bar{\rho}_s\delta(\epsilon/a-E_{21})\left[\delta_{l0}(\Delta_s-A) - \frac{T_\gamma}{T_s-T_\gamma}\left\{
\frac{\delta_{l1}}{3}(v_\gamma- v) + \sum_{l'=2}^\infty \delta_{ll'}\Theta_{l'}\right\}\right] \\
%-\frac{\delta_{l1}}{3}a\bar{\rho}_s   v \,\partial_\epsilon\left[\epsilon \delta(\epsilon/a-E_{21})\right]
-\dot{\tau}_c\left[ (\delta_{l0}-1)F_l - \frac{\delta_{l1}}{3}v \epsilon\partial_\epsilon\bar{f} + \delta_{l2}\frac{F_2}{10}\right]
-\dot{\bar{\tau}} \left[F_l+\delta_{l0}(\Delta_{HI}-\Delta_{T_s}-A)\bar{f} \right]
-\delta_{l1} \frac{v}{3}\left( \partial_\epsilon[\epsilon\dot{\bar{f}}] +\dot{\bar{\tau}}\epsilon\partial_\epsilon\bar{f}\right)
.
\label{scalarprop}
\end{multline}
In the Newtonian gauge the acceleration $A =-\psi$, the scale factor perturbation $h = -\phi$, and the shear $\sigma=0$. In the synchronous gauge $A=0$, $h = h_s/6$ and the shear $\sigma =(\dot{h}_s + 6\dot{\eta}_s)/2k$, where $h_s$ and $\eta_s$ are the usual synchronous gauge quantities.
Equation~(\ref{scalarprop}) is valid in a general gauge but the individual
terms are not gauge-invariant. In particular, under a change of
velocity field, $u^a \mapsto u^a + w^a$, the multipoles transform as
\begin{eqnarray}
F_0 &\mapsto& F_0 - \frac{w}{k}\left[\dot{\bar{f}}
+ \mathcal{H}\epsilon \partial_\epsilon \bar{f}\right] \\
F_1 &\mapsto& F_1 + \frac{1}{3} w \epsilon
\partial_\epsilon \bar{f}  \\
F_l &\mapsto & F_l \quad (l>1) .
\end{eqnarray}
For completeness, the other variables in Eq.~(\ref{scalarprop}) transform
as
\begin{eqnarray}
\dot{h} &\mapsto& \dot{h} - \frac{1}{k}(\dot{w}\clh + w \dot{\clh})
+ \frac{1}{3} kw ,  \\
\sigma &\mapsto& \sigma + w ,  \\
A &\mapsto& A + \frac{\dot{w}}{k} + \frac{\clh w}{k} ,  \\
\Delta_s &\mapsto & \Delta_s - \frac{\dot{\bar{\rho}}_s}{\bar{\rho}_s}
\frac{w}{k} , \\
v &\mapsto& v - w .
\end{eqnarray}
With these relations, one can establish the gauge-independence of
Eq.~(\ref{scalarprop}).

The line-of-sight solution is
\begin{multline}
F_l = e^{-\tau_\epsilon}\int \ud\eta e^{-\tau_c+\bar{\tau}} \biggl\{ \left[ \left\{a\bar{\rho}_s(\Delta_s-A) -a\bar{\tau}_s(\Delta_{HI}-\Delta_{T_s} - A)\bar{f} \right\}\delta(\epsilon/a-E_{21}) + \dot{h}\epsilon\partial_\epsilon \bar{f}- \dot{\tau}_cF_0 \right] j_l(k \chi) \\
-\left(\frac{k\sigma}{3} \epsilon\partial_\epsilon \bar{f}
+\dot{\tau}_c\frac{ F_2}{4}
%+a\bar{\rho}_s\delta(\epsilon/a-E_{21})\frac{\bar{T}_\gamma}{\bar{T}_s-\bar{T}_\gamma}\frac{5\Theta_2}{2}
\right)\left[ 3j_l''(k\chi) +  j_l(k\chi) \right]
- a \left[ (\bar{\rho}_s-\bar{\tau}_s\bar{f}) \partial_\epsilon \left[\epsilon\delta(\epsilon/a-E_{21})\right] v -\bar{\rho}_s \frac{\bar{T}_\gamma}{\bar{T}_s-\bar{T}_\gamma}(v-v_\gamma) \delta(\epsilon/a-E_{21})\right]j_l'(k\chi)
\\
+ (k A-\dot{\tau}_cv) \epsilon\partial_\epsilon \bar{f} j_l'(k\chi)
- a\bar{\rho}_s\delta(\epsilon/a-E_{21}) \frac{\bar{T}_\gamma}{\bar{T}_s-\bar{T}_\gamma}
%\sum_{l=2}^\infty 2^{-l}(2l+1)\Theta_l\sum_{m=0}^{l/2} \frac{(2l-2m)!}{m! (l-m)!(l-2m)!}\frac{\ud^{l-2m}}{\ud(k\chi_\epsilon)^{l-2m}} j_l(k\chi_\epsilon)
\sum_{l'=2}^\infty (2l'+1)\Theta_{l'} i^{l'} P_{l'}\left(-\frac{i}{k}\frac{\ud}{\ud\chi_\epsilon}\right)j_l(k\chi)
\biggr\}
\end{multline}
(for $l>1$), the spin temperature perturbation is
\begin{multline}
\Delta_{T_s} - \Delta_{T_\gamma} =  (R_\gamma-R_g)\delta C_{10} + C_{10}(R_g \Delta_{T_g} - R_\gamma\Delta_{T_\gamma})
+\half\tau_\eta A_{10}C_{10}\frac{\bar{T}_g-\bar{T}_\gamma}{T_\star} R_g R_\gamma
\\
\times
\left[ \Delta_{HI} - \frac{\dot{h}}{\clh} - \frac{kv}{3\clh} - A
+ \Delta_{C_{10}} + 2\Delta_{T_\gamma}(C_{10}R_\gamma-1) - 2R_\gamma \delta C_{10} + \frac{\bar{T}_\gamma}{\bar{T}_g - \bar{T}_\gamma}\left(\Delta_{T_g}-\Delta_{T_\gamma}\right)\right],
\end{multline}
and the gas temperature perturbations evolve with
\begin{equation}
\dot{\Delta}_{T_g} = - \frac{2}{3} k v -2\dot{h}
 -\frac{8 a \sigma_T \bar{\rho}_\gamma \bar{x}_e}{3 m_e c(1+f{_\text{He}}+\bar{x}_e)}\left[ \left(1-\frac{\bar{T}_\gamma}{\bar{T}_g}\right) \left\{4\Delta_{T_\gamma}-A +  \frac{\Delta_{x_e}}{1+\bar{x}_e/(1+\fHe)}\right\}
 + \frac{\bar{T}_\gamma}{\bar{T}_g}(\Delta_{T_g} -\Delta_{T_\gamma}) \right],
\end{equation}
where we again neglected helium fraction perturbations. Equation~\eqref{sound_speed} in the main text holds in any gauge.

We choose to work in the synchronous gauge, evaluating the line-of-sight solution integrated over a frequency window function. We use the synchronous gauge because this is stable for isocurvature mode evolution, and because conventional calculations invariably use the baryon (or CDM) power spectrum in the synchronous gauge as output by CMBFAST or CAMB. Since both $v-v_c$ and $\tau_\eta$ are small on scales where our calculation is applicable (much larger than the line width), in the synchronous gauge ($v_c=0$) terms involving $\tau_\eta v$ can be dropped to good accuracy.

We define a window function $W_f(\epsilon)$ so that we observe $\int \ud \epsilon W_f(\epsilon) f$. Then integrating over energies, assuming the window function is much broader than the line width, we use the functions
\begin{equation}
G(\eta) \equiv \int \ud \epsilon e^{\bar{\tau}-\tau_\epsilon}W_f(\epsilon) a\bar{\rho}_s \delta(\epsilon/a-E_{21}) = \frac{1-e^{-\tau_\eta}}{\tau_\eta} a^2\bar{\rho}_s W_f(\epsilon_\eta)
\end{equation}
\begin{equation}
G_\tau(\eta) \equiv \int \ud \epsilon e^{\bar{\tau}-\tau_\epsilon}W_f(\epsilon) a\bar{\tau}_s \bar{f} \delta(\epsilon/a-E_{21}) = \left[1-\frac{\tau_\eta e^{-\tau_\eta}}{1-e^{-\tau_\eta}}\right] G(\eta)
\end{equation}
\begin{equation}
V(\eta) \equiv \int \ud \epsilon \,e^{\bar{\tau}-\tau_\epsilon} W_f(\epsilon) \epsilon\partial_\epsilon \bar{f} = -  e^{-\tau_{\eta}}\frac{a^2\bar{\rho}_s}{\clh} W_f(\epsilon_\eta) + \int^\eta \ud \eta' a(\eta')\frac{\partial}{\partial \eta'}\left(\frac{1-e^{-\tau_{\eta'}}}{\tau_{\eta'}}\frac{a(\eta') \bar{\rho}_s(\eta')}{\clh(\eta')}\right)W_f(\epsilon_{\eta'}),
\end{equation}
where $W_f(\epsilon) = -W_f(z)E_{21}/\epsilon^2 = W_f(a)/E_{21}$.
For 21cm we assume we measure some averaged brightness temperature
\begin{equation}
\bar{T}_b = \int \ud\nu W_T(\nu) T_b(\nu).
\end{equation}
Then
\begin{equation}
\bar{T}_b = \int \ud a W_T(a)  \frac{c h_p^3}{2k_B} \epsilon f
\end{equation}
where $W_T(a)$ is a window over dimensionless frequency, so
\begin{equation}
W_f(\epsilon) = \frac{c h_p^3}{2 k_B} a W_T(a).
\end{equation}

\section{Evaluation of $\mathcal{N}_{\nu\,0}$}
\label{app:deltan}

The contribution from radiative transitions in the 21cm line to the
evolution of the spin temperature is
\begin{eqnarray}
\frac{\partial \beta_s}{\partial \tau} &=& \beta_* \left(1+\frac{N_1}{N_0}
\right) \left[(1+\mathcal{N}_{\nu\,0})A_{10} - 3 \frac{N_0}{N_1}A_{10}
\mathcal{N}_{\nu\,0}\right] \nonumber \\
&\approx& 4\beta_* A_{10} \left(1-\frac{\beta_s}{\beta_*}\mathcal{N}_{\nu\,0}
\right) .
\end{eqnarray}
Here, recall, $\mathcal{N}_{\nu\,0}$ is the isotropic part (in the gas
rest frame) of the photon occupation number integrated over the line
profile. The occupation number has contributions from the background and
perturbed CMB and the 21-cm line radiation, i.e.
\begin{equation}
\mathcal{N}_\nu = \frac{k_B \bar{T}_\gamma}{E}(1+\Theta) + \frac{h_p^3}{2}
(\bar{f} + \delta f) ,
\end{equation}
making the Rayleigh-Jeans approximation for the CMB. Evaluating
\begin{eqnarray}
\mathcal{N}_{\nu\,0} &=& \frac{1}{4\pi} \int \ud \tilde{E} \ud \tilde{\Omega}\,
\mathcal{N}(\tilde{E},\tilde{\ve}) \Phi(\tilde{E}-E_{21}) \nonumber \\
&=& \frac{1}{4\pi} \int \ud E \ud \Omega \, \frac{E}{\tilde{E}}
\mathcal{N}(E,\ve) \Phi(\tilde{E}-E_{21})
\end{eqnarray}
for the CMB, we find
\begin{equation}
\mathcal{N}_{\nu\,0}^{(\mathrm{CMB})} = \frac{\bar{T}_\gamma}{T_*}(1 + \Delta_{T_{\gamma}}) .
\end{equation}

The 21cm part takes more work. First consider the contribution from
$\delta f$; since $\delta f$ is first-order, we only require its monopole
in the conformal Newtonian gauge. As we are integrating over the line profile, at any time
$\eta$ we are
including only 21-cm radiation that was `produced' at earlier times
within a narrow time window $\Delta \eta \sim \Delta E / (\clh E_{21})$
where $\Delta E$ is the line width. We therefore
generalize Eq.~(\ref{result}) for $\epsilon$ very close
to $\epsilon_\eta$ under the assumption that the perturbations have
(comoving) wavelength $\gg \Delta \eta$. Extracting the monopole, we find
\begin{equation}
\frac{h_p^3}{2} \delta f_0(\eta,\vx,\epsilon) \approx
\left(\frac{\bar{T}_s - \bar{T}_\gamma}{T_*}\right)_\epsilon
\left[\frac{1}{\clh}\left(\dot{\phi}-\frac{1}{3}\vgrad \cdot
\vv\right)\bar{\tau}e^{-\bar{\tau}} + (\Delta_s+\psi)\left(1-e^{-\bar{\tau}}\right)
-(\Delta_{H_I} - \Delta_{T_s} + \psi)\left[1-(1+\bar{\tau})e^{-\bar{\tau}}\right]
\right] .
\end{equation}
The divergence of the gas velocity arises here from the monopole of the
redshift-space distortion:
\begin{equation}
\frac{1}{4\pi} \int \ud \Omega_\ve \, \ve \cdot (\ve \cdot \vgrad \vv)
= \frac{1}{3} \vgrad \cdot \vv .
\end{equation}
Integrating over the line profile, we obtain the following contribution to
$\cln_{\nu 0}$ from 21cm perturbations:
\begin{eqnarray}
\mathcal{N}_{\nu\,0}^{(\delta f)} &=& \left(\frac{\bar{T}_s - \bar{T}_\gamma}{T_*}\right)
\left[(\Delta_s+\psi)\left(1-\frac{1-e^{-\tau_\eta}}{\tau_\eta}\right)
-(\Delta_{HI}-\Delta_{T_s}+\psi)\left(1+e^{-\tau_\eta} - 2 \frac{1-e^{-\tau_\eta}}{\tau_\eta}\right) \right. \nonumber \\
&&\mbox{} \left.
+ \frac{1}{\clh}\left(\dot{\phi}-\frac{1}{3}\vgrad \cdot \vv\right)
\left(\frac{1-e^{-\tau_\eta}}{\tau_\eta} - e^{-\tau_\eta}\right)\right]
\end{eqnarray}
We also need the contribution
from $\bar{f}$. Correct to linear-order in $\vv$, this is simply
\begin{equation}
\mathcal{N}_{\nu\,0}^{(\bar{f})} = \left(\frac{\bar{T}_s
- \bar{T}_\gamma}{T_*}\right) \left(1-\frac{1-e^{-\tau_\eta}}{\tau_\eta}\right) ;
\end{equation}
non-linear corrections in $\vv$ may be important if $E_{21} |\vv| \gsim
\Delta E$. Combining all results, and expanding to first-order in
$\tau_\eta$, we obtain our desired result
\begin{equation}
\mathcal{N}_{\nu\,0} \approx \frac{\bar{T}_\gamma}{T_*}(1 + \Delta_{T_{\gamma}})
+ \frac{\tau_\eta}{2}\left(\frac{\bar{T}_s - \bar{T}_\gamma}{T_*}\right)
\left[1+\Delta_s+\psi+\frac{1}{\clh}\left(\dot{\phi}-\frac{1}{3}\vgrad \cdot
\vv\right)\right] .
\end{equation}
In this approximation, the perturbed Rayleigh-Jeans' brightness temperature,
$T_* \mathcal{N}_{\nu\,0}$, that we use in the main text is
\begin{equation}
T_\gamma + T_{b0} = \bar{T}_\gamma (1+\Delta_{T_\gamma})
+ \frac{\tau_\eta}{2}(\bar{T}_s - \bar{T}_\gamma)\left[1+\Delta_s+\psi+\frac{1}{\clh}\left(\dot{\phi}-\frac{1}{3}\vgrad \cdot
\vv\right)\right] .
\end{equation}
As we might have anticipated, this can be expressed in terms of the
(perturbed) optical depth
$\hat{\tau}_\eta$ of Eq.~\ref{eq:exactopd}, and the monopole of the
CMB temperature $T_\gamma$ as $T_{b0} = \hat{\tau}_\eta(T_s - T_\gamma)/2$.

\section{Ionization fraction perturbations}
\label{xepert}
For our approximate analysis of ionization fraction perturbations we start the perturbation evolution after Helium  has recombined, so we take $n_{HI}=(1-x_e)n_H$ and $n_e=x_e n_H$ where $n_H$ is the total number density of ionized and unionized hydrogen. We then use the effective equation of RECFAST~\cite{Seager:1999km}
\begin{equation}
\dot{x}_e = -a C_r \left( x_e^2 n_H \alpha - \beta(1-x_e)e^{-E_{H2s}/k_B T_g}\right)
, \label{dotx}
\end{equation}
where $E_{H2s}$ is the energy of the transition from the ground to the 2s state and
\begin{equation}
C_r \equiv \frac{1 + \clk\Lambda n_H(1-x_e)}{1+\clk(\Lambda+\beta)n_H(1-x_e)}.
\end{equation}
The recombination and photoionization coefficients are related by
\begin{equation}
\beta = \alpha \left(\frac{m_e k_B T_g}{2\pi h^2}\right)^{3/2}  e^{-E_{2s}/k_B T_g}
, \end{equation}
where $E_{2s}$ is the ionization energy from the $2s$ state, and $\alpha$ is also a function of the temperature fit by
\begin{equation}
\alpha = F   \frac{ a_\alpha (T_g/10^4 K)^b}{1+c  (T_g/10^4 K)^d} \text{m}^3\text{s}^{-1}.
\end{equation}
Here $F$ is a fudge factor taken to be $1.14$ and $a_\alpha = 4.309\times 10^{-19}$, $b=-0.6166$, $c=0.6703$, $d=0.5300$. The constant two photon $2s$--$1s$ decay rate is $\Lambda = 8.22458\rm{s}^{-1}$. Dependence on the expansion rate enters through the cosmological redshifting term $\clk$, given in the background by $\clk=a\lambda_{H2p}^3/(8\pi \clh)$.

%We use initial conditions from the Saha equation that follows enforcing cancellation in the terms on the %right hand side of Eq.~\eqref{dotx},
%\begin{equation}
%\frac{x_e^2}{1-x_e}= \frac{1}{n_H}\left(\frac{m_e k_B T_g}{2\pi h^2}\right)^{3/2} e^{-E_i/k_B T_g},
%\end{equation}
%where $E_i=E_{2s} + E_{H2s}$ is the ionization energy of hydrogen. At the beginning of recombination when %the Saha equation is valid one might think the perturbation $\Delta_{x_e}$ then satisfies
%\begin{equation}
%\Delta_{x_e} = \frac{1-\bar{x}_e}{2-\bar{x}_e}\left(-\Delta_b + \left(\frac{3}{2} + \frac{E_i}{k_B %\bar{T}_g}\right)\Delta_{T_g}\right).
%\end{equation}
% This shows that $\Delta_{x_e}$ grows rapidly once the hydrogen starts to recombine, and grows so that $\clo(\Delta_{x_e}) \sim -\clo(\Delta_b)$. The sign is opposite because overdense regions allow recombination to proceed further before freeze out.
%This will only hold for $\dot{x}_e\sim 0$, and actually does not appear to be very useful. It does however show that $\Delta_{x_e}$ is zero until recombination starts, and suggests that $\Delta_{x_e}$ can be at least the same order as $\Delta_b$.
A perturbed version of Eq.~\eqref{dotx} may be inappropriate due to the effect of perturbation velocities on the escape probabilities that went in to deriving the result. However the detailed perturbation evolution during recombination does not affect the 21cm absorption signal as long as the residual perturbations after recombination are correct, at which point recombination is limited solely by the low rate of electron capture to an excited state. A full analysis is beyond the scope of this paper, so we proceed from the perturbed version of Eq.~\eqref{dotx} and argue this is sufficient:
\begin{multline}
\dot{\Delta}_{x_e} = \frac{\dot{x}_e}{x_e}\left(\Delta_{C_r} + \Delta_\alpha - \Delta_{x_e} - A\right) \\-
a C_r\left\{ (2\Delta_{x_e} + \Delta_H) \alpha x_e n_H +  \left\{\Delta_{x_e} - \left(\frac{3}{2}+\frac{E_{2s}}{k_B T_g}\right)\left(\frac{1}{x_e}-1\right)\Delta_{T_g}\right\}\beta e^{-E_{H2s}/k_B T_g}\right\}.
\label{delta_xe}
\end{multline}
We generalize $\clk$ to a perturbed universe as $\clk = 3\lambda_{H2p}^3/(8\pi \nabla_a u_g^a)$ to account for the local baryon expansion rate; this should be correct on scales sufficiently large compared to the mean Lyman-$\alpha$ photon interaction length (which is very short). The perturbed terms are then
\begin{eqnarray}
\Delta_{C_r} &=& -\frac{\beta \clk n_H\left\{ \left[ \Delta_H + \Delta_\clk + \Delta_\beta(1+\clk\Lambda n_H (1-x_e))\right](1-x_e) - x_e \Delta_{x_e}\right\}}{(1+\clk\Lambda n_H(1-x_e))(1+\clk(\beta+\Lambda)n_H(1-x_e))} \\
\Delta_\beta &=& \Delta_\alpha + \left(\frac{3}{2}+\frac{E_{2s}}{k_B T_g}\right)\Delta_{T_g}\\
\Delta_\alpha &=& \frac{ b + c(T_g/10^4 K)^d(b-d)}{1+c(T_g/10^4 K)^d}\Delta_{T_g}\\
\Delta_\clk &=& - \frac{\dot{h}}{\clh} -\frac{k v}{3\clh} - A.
\end{eqnarray}
In the synchronous gauge the dominant term on large scales is initially due to fluctuations in the expansion rate, $\Delta_\clk \sim \dot{\Delta}_b/3\clh$:  $\Delta_{x_e}$ is the same sign as $\Delta_c$ because overdensities expand less fast and hence have a higher ionization fraction than the average
%This is the opposite to what one would expect from the perturbed Saha equation, and
($\Delta_\clk$ appears to have been ignored in e.g. Refs.~\cite{Liu:2000dy,Singh:2001ub}). At later times when $x_e \ll 1$ the main effect is from perturbations in the hydrogen density and temperature; this leads to $\Delta_{x_e}$ being the opposite sign to $\Delta_c$ at late times because overdensities recombine more efficiently. Note that at late times some terms in Eq.~\eqref{delta_xe} are invalid because $\Delta_{T_g}E_i/k_B T_g$ becomes of order unity, however the error is harmless because the entire term is exponentially suppressed: at late times the evolution is given approximately by
\begin{eqnarray}
\dot{x}_e &\approx & -a \alpha x_e^2 n_H\\
\dot{\Delta}_{x_e} &\approx& -a \alpha x_e n_H \left( \Delta_\alpha + \Delta_{x_e} +\Delta_{H}\right).
\end{eqnarray}
It is not necessary to model the early evolution correctly to get approximately the correct late-time answer. Indeed the late-time evolution neglecting velocity effects should be quite accurate as recombination is limited by the low electron capture probability. However the overall ionization fraction evolution is limited by the precision of the RECFAST model, in which a single fudge-factor accounts for deviations of an effective three level atom model from the full result.

Using the equations given here the effect on the 21cm power spectrum of neglecting $\Delta_{x_e}$ is ${\clo (2\%)}$ at $z\sim 50$, almost entirely due to the indirect effect on the evolution of the temperature perturbation. Note that, at the Jeans' scale, ionization fraction perturbations also have a linear effect on the baryon density evolution due to the modified evolution of the gas temperature perturbation (and hence the baryon pressure perturbation).

\section{Spherical Bessel function integrals and approximations}
\label{Bessels}
Integrals of products of spherical Bessel functions can be done analytically using the result~\cite{Gradshteyn00}
\begin{equation}
\int_0^\infty \ud r j_l(r) j_{l'}(r) r^{-n} = \frac{ \pi \Gamma\left(n+1\right)\Gamma\left(\frac{l+l'+1-n}{2}\right) }
{2^{n+2}\Gamma\left(\frac{l-l'+2+n}{2}\right)\Gamma\left(\frac{l+l'+3+n}{2}\right)\Gamma\left(\frac{l'-l+2+n}{2}\right) }.
\label{bess_int}
\end{equation}
In particular, using this result in combination with the spherical Bessel equation we have
\begin{eqnarray}
l(l+1)\int_{-\infty}^{\infty} \ud \ln r\, [j_l(r)]^2 &=& \frac{1}{2} \\
l(l+1)\int_{-\infty}^{\infty} \ud \ln r\, j_l(r) j_l''(r) &=& -\frac{1}{6} \\
l(l+1)\int_{-\infty}^{\infty} \ud \ln r\, [j_l''(r)]^2 &=& \frac{1}{10} + \frac{4}{15(l-2)(l+3)}\approx \frac{1}{10},
\end{eqnarray}
where the approximation holds for high $l$ of most interest in this paper.

At high $l$ the Bessel functions oscillate very rapidly compared to the scale of variation in the power spectra. Direct numerical integration becomes slow, but quite accurate results can be obtained by averaging over the oscillations.
Defining $\nu \equiv l+1/2$, for $r^2/\nu^2 \agt 1+ \nu^{-2/3}$ we can use the approximation~\cite{Gradshteyn00}
\begin{equation}
%j_l(r) = \frac{\cos(b)\cos(-v (\tan(b)-b)+1/4*\pi)}{v\sqrt{\sin(b)})}
j_l(r) \approx \frac{\sin \left( \sqrt{r^2-\nu^2} - \arccos(\nu/r)/\nu + \pi/4\right)}{r(1-\nu^2/r^2)^{1/4}}.
\end{equation}
Neglecting oscillatory parts that closely average to zero we then have for $l \gg 1$
\begin{eqnarray}
\la [j_l(r)]^2 \ra &\sim& \frac{1}{2r\sqrt{r^2-\nu^2}} \\
\la [j_l''(r)]^2 \ra &\sim& \frac{(r^2-\nu^2)^{3/2}}{2r^5} \\
\la j_l''(r) j_l(r) \ra &\sim& -\frac{\sqrt{r^2-\nu^2}}{2r^3}.
\end{eqnarray}

\section{Nonlinear CDM power spectra}
\label{nonlinear}
At a given redshift the two leading corrections to the power spectrum are given in terms of the linear-theory matter power spectrum at that redshift, $\clp(k)\equiv k^3P(k)/(2\pi^2)$, by~\cite{Vishniac83,Suto91,Makino92,Scoccimarro:2004tg}
\begin{equation}
\clp_{13}(k)=\int_0^\infty \ud r I_{13}(k,r)\equiv\frac{\clp(k)}{504}\int_{-\infty}^\infty \ud \ln r\,
\clp(k r)
\left[\frac{12}{r^4}-\frac{158}{r^2}+100  -42 r^2 +\frac{3}{r^5}\left(r^2-1\right)^3
(7 r^2 +2)\ln\left|\frac{1+r}{1-r}\right|\right]
\label{P13}
\end{equation}
and
\begin{equation}
\clp_{22}(k)=\int_0^\infty \ud r \int_{-1}^1 \ud x  I_{22}(k,r,x)\equiv\frac{1}{196}\int_{-\infty}^\infty \ud \ln r\, \clp\left(k r\right)
\int_{-1}^1 \ud x \clp\left(k\left[1+r^2-2 r x\right]^{1/2}\right)
\frac{\left(3 r +7 x-10 r x^2\right)^2}{r^2\left(1+r^2-2 r x\right)^{7/2}}~.
\label{P22}
\end{equation}
For the small scales of interest to us here the dominant contribution comes from $q \alt k$, with significant mode coupling to all scales where the spectrum is growing logarithmically. This can become numerically difficult because for $r \ll 1$ the two terms $P_{22}$ and $P_{13}$ become large but almost cancel. For $r<r_s$ we therefore use an approximate series expansion, switching to the full result at $r>r_s$. The expression for the correction to the matter power spectrum can then be written
\begin{multline}
\label{intsplit}
\clp_{13}(k)+\clp_{22}(k) \approx \left[\frac{8126}{2205}  - \frac{22}{21} \frac{\ud \ln \clp}{\ud \ln k} + \frac{1}{10}\left\{ \left(\frac{\ud \ln \clp}{\ud \ln k} \right)^2 + \frac{\ud^2 \ln \clp}{\ud (\ln k)^2}   \right\}\right]  \clp(k)\int_{-\infty}^{\ln r_s} \ud \ln r \, \clp(k r)
\\
 +  \int_{r_s}^\infty \ud r I_{13}(k,r)
+ \left[ 2\int_{r_s}^\epsilon \ud r \int_{-1}^1 \ud x
+  \int_\epsilon^{1-\epsilon} \ud r  \int_{-1}^1 \ud x
+  \int_{1-\epsilon}^{1+\epsilon} \ud r  \int_{-1}^{(1+r^2-\epsilon^2)/2r} \ud x
+  \int_{1+\epsilon}^\infty \ud r \int_{-1}^1 \ud x\right]I_{22}(k,r,x)
\end{multline}
%\begin{multline}
%P_{13}(k)+P_{22}(k) \approx \left[\frac{5038}{2205}  - \frac{94}{105} \frac{\ud \ln P}{\ud \ln k} + \frac{1}{5}\left\{ %\left(\frac{\ud \ln P}{\ud \ln k} \right)^2 + \frac{\ud^2 \ln P}{\ud (\ln k)^2}   \right\}\right] \frac{k^3 %P(k)}{(2\pi)^2}\int_0^{r_s} \ud \ln r \, r^3 P(k r)
%\\
% +  \int_{r_s}^\infty \ud r I_{13}(k,r)
%+ \left[ 2\int_{r_s}^\epsilon \ud r \int_{-1}^1 \ud x
%+  \int_\epsilon^{1-\epsilon} \ud r  \int_{-1}^1 \ud x
%+  \int_{1-\epsilon}^{1+\epsilon} \ud r  \int_{-1}^{(1+r^2-\epsilon^2)/2r} \ud x
%+  \int_{1+\epsilon}^\infty \ud r \int_{-1}^1 \ud x\right]I_{22}(k,r,x)
%\end{multline}
where $r_s < \epsilon \le 1/2$ (numerically it is better to chose $\epsilon \sim 1/2$). For a scale invariant primordial spectrum the small scale power spectrum is $\clp(kr) \sim \clp(k)[1 - \ln(r)/\ln(r_0)]^2$, where $r_0$ corresponds to the much larger scale $k_0$ where logarithmic growth starts. The main non-linear contribution to the power spectrum can then be approximated from the first term in Eq.~\eqref{intsplit} as $\sim - \ln(r_0)[\clp(k)]^2\sim \ln(k/k_0)[\clp(k)]^2$. On the Jeans' scale at $z=50$ this implies a fractional second-order contribution to the power spectrum of $\sim \clo(10)\clp(k)$: on small scales non-linear effects are more important than one might naively think~\cite{Jain:1993jh}. See Fig~\ref{nonlinpk} for typical numerical results.

The 21cm angular power spectrum on small scales is also sensitive to redshift distortions, for which we need to estimate the second-order velocity power spectrum and the cross-correlation with the density. The results are similar to those for the density with
\begin{equation}
\clp_{13}^{v v}(k)= \frac{\clp(k)}{168}\int_{-\infty}^\infty \ud \ln r\,
\clp(k r)
\left[\frac{12}{r^4}-\frac{82}{r^2}+4  -6 r^2 +\frac{3}{r^5}\left(r^2-1\right)^3
(r^2 +2)\ln\left|\frac{1+r}{1-r}\right|\right],
\label{Pv13}
\end{equation}
and
\begin{equation}
\clp_{22}^{v v}(k)=\frac{1}{196}\int_{-\infty}^\infty \ud \ln r\, \clp\left(k r\right)
\int_{-1}^1 \ud x\, \clp\left(k\left[1+r^2-2 r x\right]^{1/2}\right)
\frac{\left( r -7 x+6 r x^2\right)^2}{r^2\left(1+r^2-2 r x\right)^{7/2}},
\label{Pv22}
\end{equation}
where $\clp^{vv}$ is the power spectrum of $k v/\clh$. The series result for use at high $l$ is
\begin{equation}
\clp_{13}^{v v}(k)+\clp_{22}^{v v}(k) \approx \left[\frac{558}{245}  - \frac{94}{105} \frac{\ud \ln \clp}{\ud \ln k} + \frac{1}{10}\left\{ \left(\frac{\ud \ln \clp}{\ud \ln k} \right)^2 + \frac{\ud^2 \ln \clp}{\ud (\ln k)^2}   \right\}\right]  \clp(k)\int_{-\infty}^{\ln r_s} \ud \ln r \, \clp(k r).
\end{equation}
Similarly the cross-correlation power spectrum is given by
\begin{equation}
\clp_{13}^{v\delta}(k)= - \frac{\clp(k)}{504}\int_{-\infty}^\infty \ud \ln r\,
\clp(k r)
\left[\frac{24}{r^4}-\frac{202}{r^2}+56  -30 r^2 +\frac{3}{r^5}\left(r^2-1\right)^3
(5r^2 +4)\ln\left|\frac{1+r}{1-r}\right|\right]
\label{Pvd13}
\end{equation}
and
\begin{equation}
\clp_{22}^{v\delta}(k)=-\frac{1}{196}\int_{-\infty}^\infty \ud \ln r\, \clp\left(k r\right)
\int_{-1}^1 \ud x\, \clp\left(k\left[1+r^2-2 r x\right]^{1/2}\right)
\frac{\left( r -7 x+6 r x^2\right)\left(-3 r -7 x+10 r x^2\right)}{r^2\left(1+r^2-2 r x\right)^{7/2}},
\label{Pvd22}
\end{equation}
with the series result
\begin{equation}
\clp_{13}^{v\delta}(k)+\clp^{v\delta}_{22}(k) \approx - \left[\frac{6382}{2205}  - \frac{34}{35} \frac{\ud \ln \clp}{\ud \ln k} + \frac{1}{10}\left\{ \left(\frac{\ud \ln \clp}{\ud \ln k} \right)^2 + \frac{\ud^2 \ln \clp}{\ud (\ln k)^2}   \right\}\right]  \clp(k)\int_{-\infty}^{\ln r_s} \ud \ln r \, \clp(k r).
\end{equation}

\bibliography{../antony,../cosmomc}

\end{document}